\documentclass[floatfix,twocolumn,showpacs,floats,aps,pre,showpacs,amsfonts,amssymb]{revtex4}
\usepackage{amsmath}
\usepackage{bm}
\usepackage{color}
\usepackage{hyphenat}
\usepackage{graphicx}
\usepackage{dcolumn}
\usepackage{esvect}
\usepackage{dcolumn}
\usepackage{hyperref}

\newlength{\sepmod}
\newcommand{\ds}{\displaystyle}

\def\eqref#1{(\ref{#1})}
\def\l{\langle}

\def\r{\rangle}

\def\d{\mathrm{d}}
\def\o{{\cal O}}
\newcommand{\prt}[2]{\ensuremath{\frac{\partial #1}{\partial #2}}}
\newcolumntype{t}[1]{D{.}{.}{#1}}

\begin{document}
\title{Error estimation and reduction with cross correlations}

\author{Martin Weigel}
\email{weigel@uni-mainz.de}
\affiliation{Theoretische Physik, Universit\"at des Saarlandes, D-66041 Saarbr\"ucken, Germany}
\affiliation{Institut f\"ur Physik, Johannes Gutenberg-Universit\"at Mainz,
  Staudinger Weg 7, D-55099 Mainz, Germany}

\author{Wolfhard Janke}
\email{janke@itp.uni-leipzig.de}
\affiliation{Institut f\"ur Theoretische Physik and Centre for Theoretical Sciences (NTZ),
  Universit\"at Leipzig, Postfach 100\,920, D-04009 Leipzig, Germany}

\date{\today}

\begin{abstract}
  Besides the well-known effect of autocorrelations in time series of Monte Carlo
  simulation data resulting from the underlying Markov process, using the same data
  pool for computing various estimates entails additional cross correlations. This
  effect, if not properly taken into account, leads to systematically wrong error
  estimates for combined quantities. Using a straightforward recipe of data analysis
  employing the jackknife or similar resampling techniques, such problems can be
  avoided. In addition, a covariance analysis allows for the formulation of optimal
  estimators with often significantly reduced variance as compared to more
  conventional averages.
\end{abstract}

%05.10.Ln  Monte Carlo methods
%75.50.Lk Spin glasses and other random magnets
%64.60.F- Equilibrium properties near critical points, critical exponents
%05.10.Cc Renormalization group methods
%02.60.Pn Numerical optimization
%75.10.Hk Classical spin models
%05.70.Fh  Phase transitions: general studies 

\maketitle

\section{Introduction}

Monte Carlo simulations, and in particular Markov chain based methods, have matured
over the last decades into a highly versatile and powerful toolbox for studies of
systems in statistical and condensed-matter physics \cite{binder:book2,berg:04},
ranging from classical spin models \cite{wj:chem} over soft-matter problems
\cite{holm:05} to quantum systems \cite{vojta:03b}. Their competitiveness with other
approaches such as, e.g., field-theoretic expansions for the study of critical
phenomena \cite{zinn-justin,kleinert:01}, is largely based on the development and
refinement of a number of advanced simulation techniques such as cluster algorithms
\cite{kandel:91a} and generalized-ensemble methods \cite{berg:92b,wang:01a}.

Equally important to the generation of simulation data, however, is their correct and
optimal analysis. In this field, a number of important advances over the techniques
used in the early days have been achieved as well. These include, e.g., the
finite-size scaling (FSS) approach \cite{barber:domb}, turning the limitation of
simulational methods to finite system sizes into a systematic tool for accessing the
thermodynamic limit, reweighting techniques \cite{ferrenberg:88a}, lifting the
limitation of numerical techniques to the study of single points in parameter space
to allow for continuous functions of estimates to be studied, as well as advanced
statistical tools such as the jackknife and other resampling schemes of data analysis
\cite{efron:book}.

Of these techniques, the statistical data analysis appears to have received the least
attention. Hence, while FSS analyses, even including correction terms, are quite
standard in computer simulation studies \cite{binder:book2}, a proper analysis and
reduction of statistical errors and bias appears to be much less common. Here,
resampling methods turn out to be very valuable. Although such techniques offer a
number of benefits over more traditional approaches of error estimation, their
adoption by practitioners in the field of computer simulations has not yet been as
universal as desirable. It is our understanding that this is, in part, due to a
certain lack in broadly accessible presentations of the basic ideas which are, in
fact, very simple and easy to implement in computer codes, as is demonstrated below.

More specifically, data generated by a Monte Carlo (MC) simulation are subject to two
types of correlation phenomena, namely (a) {\em autocorrelations\/} or temporal
correlations for the case of Markov chain MC (MCMC) simulations, which are directly
related to the Markovian nature of the underlying stochastic process and lead to an
effective reduction of the number of independently sampled events and (b) {\em cross
  correlations\/} between different estimates extracted from the same set of original
time series coming about by the origin of estimates in the same statistical data
pool. The former can be most conveniently taken into account by a determination of
the relevant autocorrelation times and a blocking or binning transformation resulting
in an effectively uncorrelated auxiliary time series \cite{flyvbjerg:89a}. Such
analyses are by now standard at least in seriously conducted simulational studies. On
the contrary, the effects of cross correlations have been mostly neglected to date
(see, however, Refs.~\cite{hammersley:64,ballesteros:96a,janke:97,weigel:00b}), but
are only systematically being discussed following our recent suggestion
\cite{weigel:09,fernandez:09}. In this article, we show how such cross correlations
lead to systematically wrong estimates of statistical errors of averaged or otherwise
combined quantities when a na\"{\i}ve analysis is employed, and how a statistically
correct analysis can be easily achieved within the framework of the jackknife
method. Furthermore, one can even take benefit from the presence of such correlation
effects for significantly reducing the variance of estimates without substantial
additional effort. We demonstrate the practical relevance of these considerations for
a finite-size scaling study of the Ising model in two and three dimensions.

The rest of this article is organized as follows. In Sec.\ II we give a general
recipe for a failsafe way of Monte Carlo data analysis, taking into account the
effects of autocorrelations and cross correlations mentioned above. After discussing
the complications for the more conventional analysis schemes (but not the jackknife
method) introduced by histogram reweighting and generalized-ensemble simulation
techniques in Sec.\ III, we outline the role of cross correlations in the process of
averaging over a set of MC estimates in Sec. IV and discuss the choice of an optimal
averaging procedure. In Sec.\ V, these ideas are applied to a simulational study of
the critical points of the two- and three-dimensional Ising models. Finally, Sec.\ VI
contains our conclusions.

\section{Monte Carlo error analysis}

Compared to the task of estimating the uncertainty in the result of a lab experiment
by simply repeating it several times, there are a number of complications in
correctly determining --- and possibly even reducing --- statistical fluctuations in
parameter estimates extracted from MCMC simulations. Firstly, due to the memory of
the Markovian process, subsequent measurements in the time series are correlated,
such that the fluctuations generically appear smaller than they are. This issue can
be resolved by a {\em blocking\/} of the original time-series data. Secondly, one
often needs to know the precision of parameter estimates which are complicated (and
sometimes non-parametric) functions of the measured observables. Such problems are
readily solved using resampling techniques such as the {\em jackknife\/}.

\subsection{Autocorrelations\label{sec:autocorr}}

Consider a general Monte Carlo simulation with the possible values $O$ of a given
observable $\o$ appearing according to a probability distribution $p(O)$. This form,
of course, implies that the system is in thermal equilibrium, i.e., that the
underlying stochastic process is stationary.
%Although, in
%practise, this condition might be quite difficult to ascertain, especially for
%systems with complex free-energy landscapes, no meaningful information about
%equilibrium properties can be extracted without it.
The probability density $p(O)$ could be identical to the Boltzmann distribution of
equilibrium thermodynamics as for the importance-sampling technique
\cite{metropolis:53a}, but different situations are conceivable as well, see the
discussion in Sec.~\ref{sec:histo} below. If we assume ergodicity of the chain, the
average
$$
\bar{O} \equiv \frac{1}{N}\sum_{i=1}^N O_i
$$
for a time series $\{O_1, O_2, \ldots\}$ of $N$ measurements is an unbiased estimator
of the mean
$$
\l O\r \equiv \int\d O\,p(O)O.
$$
In contrast to $\l O\r$, the estimator $\bar{O}$ is a random number, which only
coincides with $\l O\r$ in the limit $N\rightarrow\infty$. Under these circumstances,
simulational results are only meaningful if in addition to the average $\bar{O}$ we
can also present an estimate of its variance $\sigma^2(\bar{O})$. Note that, although
the distribution $p(O)$ of individual measurements might be arbitrary, by virtue of
the central limit theorem the distribution of the averages $\bar{O}$ must become
Gaussian for $N\rightarrow\infty$. Hence, the variance $\sigma^2(\bar{O})$ is the
(only) relevant parameter describing the fluctuations of $\bar{O}$. If subsequent
measurements $O_1$, $O_2$, $\ldots$ are uncorrelated, we have
\begin{equation}
  \label{eq:sigma_noautocorr}
  \sigma^2(\bar{O}) \equiv \l\bar{O}^2\r-\l\bar{O}\r^2 = \frac{\sigma^2(O)}{N},  
\end{equation}
which can be estimated without bias from \cite{brandt:book}
\begin{equation}
  \label{eq:variance_of_mean_notautocorr}
  \hat{\sigma}^2(\bar{O}) = \frac{1}{N(N-1)}\sum_{i=1}^N(O_i-\bar{O})^2,
\end{equation}
i.e., $\langle \hat{\sigma}^2(\bar{O}) \rangle = \sigma^2(\bar{O})$. This is what we
do when estimating the statistical fluctuations from a series of independent lab
experiments. Markov chain simulations entail the presence of temporal correlations,
however, such that the connected autocorrelation function,
\begin{equation}
  \label{eq:autocorrelation_function}
  C_O(s,t) \equiv \l O_s O_t\r-\l O_s\r\l O_t\r  
\end{equation}
is non-zero in general (see, e.g., Ref.~\cite{sokal:97}). Stationarity
of the chain implies that $C_O(s,s+t) = C_O(0,t) \equiv C_O(t)$. Then, the variance
of $\bar{O}$ becomes
\begin{equation}
  \label{eq:sigma_autocorr}
  \sigma^2(\bar{O}) = \frac{\sigma^2(O)}{N}
  \left[1+2\sum_{t=1}^N\left(1-\frac{t}{N}\right)\frac{C_O(t)}{C_O(0)}\right].
\end{equation}
Monte Carlo correlations decline exponentially, i.e.,
\begin{equation}
  \label{eq:exponential_autocorr}
  C_O(t)\sim C_O(0)e^{-t/\tau_\mathrm{exp}(O)}  
\end{equation}
to leading order, defining the {\em exponential autocorrelation time\/}
$\tau_\mathrm{exp}(O)$. Due to this exponential decay, for $N\gg \tau_\mathrm{exp}$
the deviations of the factors $1-t/N$ of Eq.~(\ref{eq:sigma_autocorr}) from unity can
be neglected \cite{priestley:book}, and defining the {\em integrated autocorrelation
  time\/} as
\begin{equation}
  \label{eq:tau_int}
  \tau_\mathrm{int}(O) \equiv \frac{1}{2}+\sum_{t=1}^N\frac{C_O(t)}{C_O(0)},
\end{equation}
one has
\begin{equation}
  \label{eq:tau_reduction}
  \sigma^2(\bar{O}) \approx \frac{\sigma^2(O)}{N/2\tau_\mathrm{int}(O)}.
\end{equation}
In view of the $1/N$ reduction of variance of the average $\bar{O}$ relative to a
single measurement in Eq.~(\ref{eq:sigma_noautocorr}), Eq.~(\ref{eq:tau_reduction})
states that the {\em effective\/} number of independent measurements in the presence
of autocorrelations is reduced by a factor of $1/2\tau_\mathrm{int}(O)$. The
autocorrelation times $\tau_\mathrm{exp}$ and $\tau_\mathrm{int}$ are not identical,
but one can show that the latter is a lower bound of the former,
$\tau_\mathrm{int}(O)\le \tau_\mathrm{exp}(O)$ \cite{sokal:89a}.

\begin{figure}[t]
  \centering
  \includegraphics[keepaspectratio=true,width=7.5cm]{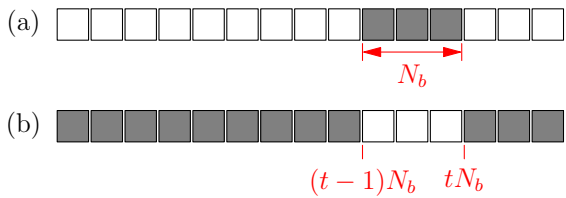}
  \caption
  {(Color online)
    Blocking transformation on a time series. In the
    binning analysis, the series is divided into blocks of length $N_b$ (a). In the
    jackknifing analysis, the blocks consist of the whole series {\em apart\/} from
    the entries of a single block (b).
    \label{fig:block}
  }
\end{figure}

As long as the autocorrelation time is finite, the distribution of averages still
becomes Gaussian asymptotically, such that for $N\gg\tau$ the variance remains the
relevant quantity describing fluctuations. To practically determine
$\sigma^2(\bar{O})$ from Eq.~(\ref{eq:sigma_autocorr}), an estimate for the
autocorrelation function is required. This can be found from the definition
(\ref{eq:autocorrelation_function}) by replacing expectation values with time
averages. It turns out, however, than upon summing over the contributions of the
autocorrelation function for different time lags $t$ in Eq.~(\ref{eq:sigma_autocorr})
divergent fluctuations are incurred, enforcing the introduction of a cut-off time
\cite{binder:domb,madras:88a}. Several approximation schemes have been developed
using such estimators, but they turn out to have severe drawbacks in being
computationally expensive, hard to automatize and in that estimating their
statistical accuracy is tedious (see Ref.~\cite{flyvbjerg:89a}).

\begin{figure}[t]
  \centering
  \includegraphics[keepaspectratio=true,scale=0.75,trim=45 48 75 78]{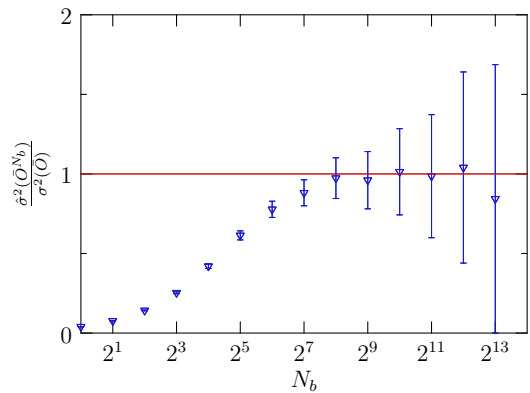}
  \caption
  {(Color online)
    Schematic representation of the estimate $\hat{\sigma}^2(\bar{O})$ of the
    variance of the average according to Eq.~(\ref{eq:variance_of_mean_notautocorr})
    for a re-blocked time series as a function of the block length $N_b$.
    \label{fig:bins}
  }
\end{figure}

A more efficient and very intuitive technique for dealing with autocorrelations
results from a blocking transformation in the spirit of the renormalization group
\cite{flyvbjerg:89a} (in fact, this idea was already formulated by Wilson
  \cite{wilson:80}). Much like block spins are defined there, one combines $N_b =
N/n$ adjacent entries of the time series,
\begin{equation}
  \label{eq:block_definition}
  {\frak B}_t := \{(t-1)N_b+1,\ldots,tN_b\},
\end{equation}
and defines block averages
\begin{equation}
  \label{eq:block_averages}
  O^{N_b}_t = \frac{1}{N_b}\sum_{k\in{\frak B}_t} O_k,\;\;\;t=1,\ldots,n,  
\end{equation}
cf.\ Fig.~\ref{fig:block}(a). This procedure results in a shorter effective time
series $\{O^{N_b}_1, O^{N_b}_2, \ldots\}$ with $n$ entries. (We assume for simplicity
that $N$ is an integer multiple of $n$.)  Obviously, the average $\bar{O}$ and its
variance $\sigma^2(\bar{O})$ are invariant under this transformation. Under the
exponential decay (\ref{eq:exponential_autocorr}) of autocorrelations of the original
series it is clear (and can be shown explicitly \cite{flyvbjerg:89a}), however, that
subsequent block averages $O^{N_b}_t$, $O^{N_b}_{t+1}$ are less correlated than the
original measurements $O_t$ and $O_{t+1}$. Furthermore, the remaining correlations
must shrink as the block length $N_b$ is increased, such that asymptotically for
$N_b\rightarrow\infty$ (while still ensuring $n\gg 1$) an uncorrelated time series is
produced. Consequently, the na\"{\i}ve estimator
(\ref{eq:variance_of_mean_notautocorr}) can be legally used in this limit to
determine the variance $\sigma^2(\bar{O})$ of the average. For the finite time series
encountered in practice, a block length $N_b \gg \tau$ and $N_b\ll N$ must be
used. This is illustrated in Figure \ref{fig:bins} showing the estimate
(\ref{eq:variance_of_mean_notautocorr}) for a blocked time series with
autocorrelation time $\tau_\mathrm{int}\approx 13$ as a function of the block length
$N_b$. It approaches the true variance $\sigma^2(\bar{O})$ from below, eventually
reaching a plateau value where any remaining pre-asymptotic deviations become
negligible compared to statistical fluctuations. If the available time series is long
enough (as compared to $\tau$), it is often sufficient to simply lump the data into
as few as some hundred blocks and restrict the subsequent data analysis to those
blocks. As a rule of thumb, in practical applications it turns out that a time series
of length $N\gtrsim 10\,000\,\tau$ is required for a reliable determination of
statistical errors as well as autocorrelation times. From
Eqs.~(\ref{eq:variance_of_mean_notautocorr}) and (\ref{eq:tau_int}) it follows that
the integrated autocorrelation time can be estimated from
\begin{equation}
  \label{eq:tauint_estimate}
  \hat{\tau}_\mathrm{int}(O) = \frac{1}{2}\frac{\hat{\sigma}^2(\bar{O}^{N_b})}
  {\hat{\sigma}^2(\bar{O}^{1})}
\end{equation}
within this scheme, where $N_b$ needs to be chosen in the plateau regime of
Fig.~\ref{fig:bins}.

\subsection{Covariance and bias}

Apart from providing an estimate of $\sigma^2(\bar{O})$ for simple quantities, the
blocking procedure has the advantage of resulting in an effectively uncorrelated
auxiliary time series which can then be fed into further statistical machinery, much
of which is restricted to the case of independent variables. Resampling schemes such
as the jackknife \cite{efron:book} provide error and bias estimates also for
non-linear functions of observables without entailing truncation error or
requiring assumptions about the underlying probability distributions.

While $\sigma^2(\bar{O})$ can be directly computed from the blocked time series of
$\o$ via the estimator (\ref{eq:variance_of_mean_notautocorr}), this approach fails
for non-linear functions $f(\l A\r, \l B\r,\ldots)$ of expectation values $\l A\r$,
$\l B\r$, $\ldots$ such as, e.g., susceptibilities or cumulants. A standard approach
for such cases is the use of error propagation formulas based on Taylor expansions
\cite{brandt:book},
\begin{equation}
  \label{eq:error_propagation}
  \sigma^2[f(\l A\r,\l B\r,\ldots)]
  = \prt{f}{\l A\r}\sigma^2(A)+\prt{f}{\l B\r}\sigma^2(B)+\cdots.
\end{equation}
Apart from the truncation error resulting from the restriction to first order in the
expansion, this entails a number of further problems: if the averages $\bar{A}$,
$\bar{B}$ etc.\ are correlated due to their origin in the same simulation,
cross-correlation terms need to be included as well. Even worse, for the case of
non-parametric parameter estimates, such as determining the maximum of some quantity
by reweighting (see below) or extracting a critical exponent with a fitting
procedure, error propagation cannot be easily used at all.

Such problems are avoided by methods based on repeated sampling from the original
data pool, using the properties of these meta samples to estimate
\mbox{(co-)variance}, reduce bias etc. These are modern techniques of mathematical
statistics whose application only became feasible with the general availability of
computers \cite{efron:book}. Most straightforwardly applicable is the jackknife
procedure, where the meta samples consist of all of the original time series apart
from one data block, cf.\ Fig.~\ref{fig:block}(b). Assume that a set of simulations
resulted in a collection of time series $\{O_{k,1}, O_{k,2},\ldots O_{k,N_k}\}$,
$k=1$, $2$, $\ldots$ for different observables, system sizes, temperatures
etc. Applying the blocking procedure described above, it is straightforward to divide
the series in effectively uncorrelated blocks. It is often convenient to use the same
number of blocks $n$ for all series (e.g., 100) which can easily be arranged for by
the blocking transformation as long as $N_k/\tau_k$ is larger than some minimum value
(e.g., $10\,000$) for each simulation and observable. If then $\vv{\frak B}_t =
({\frak B}_{1,t},\ldots,{\frak B}_{k,t})^T$ denotes the $t^\mathrm{th}$ block over
all series according to Eq.~(\ref{eq:block_definition}), where for a constant number
of blocks the block lengths $N_{b,k} = N_k/n$ might vary between the different series
under consideration, one defines the corresponding {\em jackknife block\/} as the
complement
\begin{equation}
  \label{eq:jackknife_block}
  {\frak J}_{k,t} := \{1,\ldots,N_k\} \setminus {\frak B}_{k,t},
\end{equation}
cf.\ Fig.~\ref{fig:block}. Considering now an estimator $\hat{\theta}(\{\vv{O}_t\})$
for some parameter $\theta$ depending on (some or all of) the different series, we
define the corresponding estimates restricted to jackknife block $\vv{\frak J}_s$,
\begin{equation}
  \label{eq:jackknife_blocks}
  \hat{\theta}_{(s)} = \hat{\theta}[(\{O_{1,t\in {\frak J}_{1,s}}\},\ldots,\{O_{k,t\in {\frak J}_{k,s}}\})^T]. 
  \end{equation}
The variation between these estimates taken from the same original data can be used
to infer the sample variance. If one denotes the average of the jackknife block
estimators (\ref{eq:jackknife_blocks}) as
\begin{equation}
  \label{eq:jackknife_average}
  \hat{\theta}_{(\cdot)} =  \frac{1}{n}\sum_{s=1}^n \hat{\theta}_{(s)},
\end{equation}
an estimate for the sample variance of the estimator $\hat{\theta}$ is given by
\cite{efron:82}
\begin{equation}
  \label{eq:jackknife_variance}
  \hat{\sigma}^2_\mathrm{jack}(\hat{\theta}) \equiv
  \frac{n-1}{n}\sum_{s=1}^n\left[\hat{\theta}_{(s)}-\hat{\theta}_{(\cdot)}\right]^2.
\end{equation}
This is very similar to the simple estimate (\ref{eq:variance_of_mean_notautocorr})
for the variance of the average, but it comes with a different prefactor which
serves a twofold purpose: it reweights the result from the effective jackknife series
of length $n-1$ to the original length $n$ and takes care of the fact that all of the
jackknife block estimates $\hat{\theta}_{(s)}$ are strongly correlated due to them
being based on (almost) the same data. The general Eq.~(\ref{eq:jackknife_variance})
forms a conservative and at most weakly biased estimate of the true variance
\cite{efron:book}, which lacks the truncation error of schemes based on
Eq.~(\ref{eq:error_propagation}) and is applicable to non-parametric parameter
estimates.

In a slight generalization of Eq.~(\ref{eq:jackknife_variance}) it is possible to
also estimate covariances. For a number of estimators
$\hat{\theta}_i(\{\vv{O}_t\})$, $i=1$, $2$, $\ldots$, a robust jackknife
estimator of the covariance matrix is given by
\begin{equation}
  \label{eq:jackknife_covariance}
  \hat{\Gamma}^2_{ij}(\hat{\theta}) \equiv
  \frac{n-1}{n}\sum_{s=1}^n\left[\hat{\theta}_{i(s)}-\hat{\theta}_{i(\cdot)}\right]
  \left[\hat{\theta}_{j(s)}-\hat{\theta}_{j(\cdot)}\right].
\end{equation}
In a similar way the bias of estimators can be reduced, i.e., deviations between the
mean of an observable and the expectation value of some estimator that disappear with
increasing sample length. For a detailed discussion we refer the reader to
Refs.~\cite{efron:82,efron:book}.

A general procedure for the analysis of simulation data based on blocking and
jackknife techniques hence has the following form:
\begin{enumerate}
\item Decide on the number $n$ of jackknife blocks to be used. For most
  purposes, of the order of $100-500$ blocks are sufficient.
\item For each original time series recorded in a collection of simulations, examine
  the block averages (\ref{eq:block_averages}) as a function of the block length
  $N/n$. If the result for $n$ blocks is in the plateau regime of
  Fig.~\ref{fig:bins} everything is fine; otherwise, one needs to record a longer
  time series (and possibly take measurements less frequently to keep the amount of
  data manageable).
\item For each parameter to be estimated, compute the $n$ jackknife
  block estimates (\ref{eq:jackknife_blocks}) as well as the average
  (\ref{eq:jackknife_average}) and combine them to calculate the variance
  (\ref{eq:jackknife_variance}). For a number of different parameter estimates
  $\hat{\theta}_i$, the jackknife block estimates can also be used to calculate the
  covariance (\ref{eq:jackknife_covariance}).
\end{enumerate}

\section{Histograms and errors\label{sec:histo}}

An increasing number of successful Monte Carlo techniques rely on reweighting and the
use of histograms \cite{binder:book2}. This includes the (multi-)histogram method of
Refs.~\cite{ferrenberg:88a,ferrenberg:89a} as well as the plethora of generalized
ensemble techniques ranging from multicanonical simulations \cite{berg:92b} to
Wang-Landau sampling \cite{wang:01a}. Such methods are based on the fact that samples
taken from a known probability distribution can always be translated into samples
from another distribution over the same state space. Assume, for simplicity, that
states are labeled $\{s_i\}$ as appropriate for a spin system. If a sequence
$\{s_i\}_t$, $t=1$, $2$, $\ldots$ was sampled from a stationary simulation with
probability density $p_\mathrm{sim}(\{s_i\})$, an estimator for the expectation value
of the observable $\o$ relative to the {\em equilibrium\/} distribution is given by
\begin{equation}
  \label{eq:general_estimate}
  \hat{O} = \frac{\ds\sum_{t=1}^N {\cal O}(\{s_i\}_t)\frac{\ds
      p_\mathrm{eq}(\{s_i\}_t)}{\ds p_\mathrm{sim}(\{s_i\}_t)}}
  {\ds\sum_{t=1}^N \frac{\ds p_\mathrm{eq}(\{s_i\}_t)}{\ds p_\mathrm{sim}(\{s_i\}_t)}}.
\end{equation}
For a finite simulation this works as long as the sampled and the equilibrium
distributions have sufficient {\em overlap\/}, such that the sampled configurations
can be representative of the equilibrium average at hand. For simple sampling one has
$p_\mathrm{sim} = \mathrm{const}$ and hence must weight the resulting time series
with the Boltzmann factor
\begin{equation}
  \label{eq:canonical_distr}
  p_\mathrm{eq}(\{s_i\}) \equiv p_\beta(\{s_i\}) = \frac{1}{Z_\beta}e^{-\beta{\cal H}(\{s_i\})},
\end{equation}
where ${\cal H}(\{s_i\})$ denotes the energy of the configuration $\{s_i\}$ and
$Z_\beta$ is the partition function at inverse temperature $\beta = 1/k_B T$. For
importance sampling, on the other hand, $p_\mathrm{sim} = p_\mathrm{eq}$, such that
averages of time series are direct estimates of thermal expectation values. If
samples from an importance sampling simulation with $p_\mathrm{sim} = p_{\beta_0}$
should be used to estimate parameters of $p_\mathrm{eq} = p_{\beta}$,
Eq.~(\ref{eq:general_estimate}) yields the familiar (temperature) reweighting
relation
\begin{equation}
  \label{eq:temperature_reweighting}
  \hat{O}_\beta = \frac{\sum_t {\cal O}(\{s_i\}_t)e^{-(\beta-\beta_0)E_t}}
  {\sum_t e^{-(\beta-\beta_0)E_t}},
\end{equation}
where $E_t = {\cal H}(\{s_i\}_t)$. Completely analogous equations can be written
down, of course, for reweighting in parameters other than temperature. Similarly,
canonical averages at inverse temperature $\beta$ are recovered from multicanonical
simulations via using Eq.~(\ref{eq:general_estimate}) with $p_\mathrm{sim} =
p_\mathrm{muca}$ and $p_\mathrm{eq} = p_\beta$.

Reliable error estimation (as well as bias reduction, covariance estimates etc.) for
reweighted quantities is rather tedious with traditional statistical techniques such
as error propagation \cite{janke:02}. Resampling methods, on the other hand, allow
for a very straightforward and reliable way of tackling such problems \footnote{Note
  that when using the Wang-Landau method as a direct estimate of the density of
  states to be used for computing thermal expectation values, due to the
  non-Markovian nature of the algorithm there is currently no known approach of
  reliably estimating the present statistical fluctuations apart from repeating the
  whole calculation a certain number of times.}. For the jackknife approach, for
instance, one computes jackknife block estimates of the type
(\ref{eq:general_estimate}) by simply restricting the set of time series to the
$s^\mathrm{th}$ jackknife block $\vv{\frak J}_s$. With the jackknife average
(\ref{eq:jackknife_average}), e.g., the variance estimate
(\ref{eq:jackknife_variance}) with $\hat{\theta} = \hat{O}$ can be straightforwardly
computed. Similar considerations apply to covariance estimates or bias reduced
estimators \cite{efron:book}. Extremal values of thermal averages can be determined
to high precision from the continuous family of estimates
(\ref{eq:temperature_reweighting}), where error estimates again follow
straightforwardly from the jackknife prescription.

\section{Variance reduction}

Temporal correlations resulting from the Markovian nature of the sampling process
have been discussed in Sec.~\ref{sec:autocorr} above, and we assume that they have
been effectively eliminated by an appropriate binning procedure. Extracting a number
of different parameter estimates $\hat{\theta}_i$, $i=1$, $2$, $\ldots$ from the same
number of original simulations it is clear, however, that also significant {\em cross
  correlations\/} between estimates $\hat{\theta}_i$ and $\hat{\theta}_j$ can
occur. These have profound consequences for estimating statistical error and reducing
it by making the best use of the available data \cite{weigel:09}.

If a given parameter estimate $\hat{\theta}$ depends on several observables of the
underlying time series that exhibit cross correlations, this fact is {\em
  automatically\/} taken into account correctly by the jackknife error estimate
(\ref{eq:jackknife_variance}). This is in contrast to error analysis schemes based on
error propagation formulae of the type (\ref{eq:error_propagation}), where any cross
correlations must be taken into account explicitly. Insofar the outlined approach of
data analysis is failsafe. We want to go beyond that, however, in trying to {\em
  optimize\/} statistical precision of estimates from the available data. If we
attempt to estimate a parameter $\theta$, we ought to construct an estimator
$$
\hat{\theta} = {\cal F}(\{\vv{O}_t\}),
$$
which is a function of the underlying time series with the property that $\l
\hat{\theta}\r = \theta$ (at least for $n\rightarrow\infty$). Obviously, there
usually will be a large number of such functions ${\cal F}$ and it is not possible,
in general, to find the estimator $\hat{\theta}$ of minimal variance. We therefore
concentrate on the tractable case where $\hat{\theta}$ is a linear
combination of other estimators $\hat{\theta}_i$, $i=1,\ldots,k$,
\begin{equation}
  \label{eq:linear_combination}
  \hat{\theta} = \sum_{i=1}^k \alpha_i \hat{\theta}_i.  
\end{equation}
There are different possibilities to ensure the condition $\l \hat{\theta}\r =
\theta$:
\begin{enumerate}
\item All estimators have the same expectation, $\l\hat{\theta}_i\r = \theta$, and
$\sum_i \alpha_i = 1$.
\item One estimator is singled out, say $\l\hat{\theta}_1\r = \theta$, $\alpha_1 =
  1$, and the rest has vanishing expectation, $\l\hat{\theta}_i\r = 0$, $\alpha_i$
  arbitrary, $i \ge 2$.
\item More complicated situations.
\end{enumerate}
The first type describes the case that we have several different estimators for the
same quantity and want to take an average of minimum variance \cite{weigel:09}. The
second case is tailored for situations where existing symmetries allow to {\em
  construct\/} estimators with vanishing expectation whose cross correlations might
reduce variance \cite{fernandez:09}.

To optimize the analysis, the parameters $\alpha_i$ in (\ref{eq:linear_combination})
should be chosen such as to minimize the variance
\begin{equation}
  \label{eq:variance_of_average}
  \sigma^2(\hat{\theta}) = \sum_{i,j = 1}^k \alpha_i\alpha_j
  \left[\l\hat{\theta}_i\hat{\theta}_j\r
    -\l\hat{\theta}_i\r\l\hat{\theta}_j\r\right] \equiv \sum_{i,j = 1}^k 
  \alpha_i\alpha_j\Gamma_{ij}(\hat{\theta})  
\end{equation}
For case one above, we introduce a Lagrange multiplier to enforce the
constraint $\sum_i\alpha_i = 1$, and the optimal choice of $\alpha_i$ is readily
obtained as
\begin{equation}
  \label{eq:covariance_weighted}
  \alpha_i = \frac{\sum_{j=1}^k [\Gamma(\hat{\theta})^{-1}]_{ij}}
  {\sum_{i,j=1}^k [\Gamma(\hat{\theta})^{-1}]_{ij}},
\end{equation}
leading to a minimum variance of
\begin{equation}
  \label{eq:minimum_variance}
  \sigma^2(\hat{\theta}) = \frac{1}{\sum_{i,j=1}^k [\Gamma(\hat{\theta})^{-1}]_{ij}}.
\end{equation}
Very similarly, case two leads to the choice \footnote{Note that Eqs.~(4) and (5) of
  Ref.~\cite{fernandez:09} contain some mistakes, but Eq.~(7) and the implementation
  (24) are correct.}
\begin{equation}
  \label{eq:covariance_weighted_other}
  \alpha_i = -\sum_{j=2}^k [\Gamma'(\hat{\theta})^{-1}]_{ij}\Gamma(\hat{\theta})_{j1},  
\end{equation}
where $\Gamma'(\hat{\theta})$ denotes the submatrix of $[\Gamma(\hat{\theta})]_{ij}$
with $i,j\ge 2$. Since the formalism for both cases is practically identical, in the
following we will concentrate on case one.

Let us take the time to compare the {\em optimal\/} choice of weights expressed in
Eqs.~(\ref{eq:covariance_weighted}) and (\ref{eq:covariance_weighted_other}) with
that used in more traditional approaches. Ignoring the presence of cross
correlations, several parameter estimates are often combined using an error weighting
scheme only, i.e., by choosing weights
\begin{equation}
  \label{eq:error_weighted}
  \alpha_i^\mathrm{err} = \frac{1/\sigma^2(\hat{\theta}_i)}
  {\sum_{i=1}^k 1/\sigma^2(\hat{\theta}_i)}.
\end{equation}
While the more general expression (\ref{eq:covariance_weighted}) reduces to the
weights (\ref{eq:error_weighted}) in the absence of correlations, the choice
(\ref{eq:error_weighted}) is {\em not optimal\/} as soon as cross correlations are
present. Still, the resulting average $\hat{\theta}$ remains a valid estimator of the
parameter $\theta$. In contrast, the usually used variance estimate derived from the
expression
\begin{equation}
  \label{eq:uncorrelated_variance}
  \sigma^{2\,\mathrm{err}}_\mathrm{uncorr}(\hat{\theta}) = \frac{1}{\sum_{i=1}^k 1/\sigma^2(\hat{\theta}_i)}
\end{equation}
is {\em no longer even correct\/} when cross correlations come into play. As will be
seen below from the discussion of Ising model simulations in
Sec.~\ref{sec:application}, $\sigma^{2\,\mathrm{err}}_\mathrm{uncorr}(\hat{\theta})$
generically leads to underestimation of the true variance, but occasionally
over-estimates are possible as well.

The practical implementation of the described scheme of weighting and error analysis
is straightforward with the toolset outlined in the previous sections. The covariance
matrix of the estimates $\hat{\theta}_i$ is readily computed via the jackknife
expression (\ref{eq:jackknife_covariance}). This allows to estimate the optimal
weights from inserting $\hat{\Gamma}_{ij}$ in Eq.~(\ref{eq:covariance_weighted}) (or
the analogue (\ref{eq:covariance_weighted_other}) for case two) and the variance of
the resulting optimal estimator is determined from the expression
(\ref{eq:minimum_variance}). In total, the necessary analysis can be summarized as
follows:
\begin{enumerate}
\item Perform a binning analysis to see whether for a given number of
  blocks $n$ the block averages (\ref{eq:block_averages}) for all time
  series at hand are effectively uncorrelated.
\item For each parameter estimate $\hat{\theta}_i$ compute the $n$ jackknife block
  estimates (\ref{eq:jackknife_blocks}) as well as their average and estimate their
  covariance matrix from Eq.~(\ref{eq:jackknife_covariance}).
\item For those estimates $\hat{\theta}_i$ to be combined into an average
  $\hat{\theta}$, an estimate of the optimal weighting parameters $\alpha_i$ is given
  by Eq.~(\ref{eq:covariance_weighted}) with the estimate $\hat{\Gamma}_{ij}$
  calculated in the previous step. Likewise, the variance of the resulting average is
  estimated from Eq.~(\ref{eq:minimum_variance}).
\end{enumerate}

In some cases, it is necessary to already have variance estimates of intermediate
data available for properly determining the jackknife block estimates
$\hat{\theta}_{i(\cdot)}$. This typically occurs when $\hat{\theta}_i$ is a parameter
resulting from an (ideally error weighted) fit to a number of data points, such as
for the case of a critical exponent, see the discussion below in
Sec.~\ref{sec:application}. In these cases it is straightforward to iterate the
jackknifing procedure to second order by considering each jackknife block as the
initial time series of another jackknife analysis \cite{berg:92a}.

In view of the sometimes counter-intuitive results of computing weighted averages
taking cross correlations into account (see the results in Sec.~\ref{sec:application}
below), it is instructive to examine the simple case of just two different estimates
$\hat{\theta}_1$ and $\hat{\theta}_2$ in somewhat more detail. This is done in
Appendix \ref{sec:app}.

\section{Application to the Ising model\label{sec:application}}

\begin{figure}[tb]
  \centering
  \includegraphics[keepaspectratio=true,scale=0.75,trim=45 48 75 78]{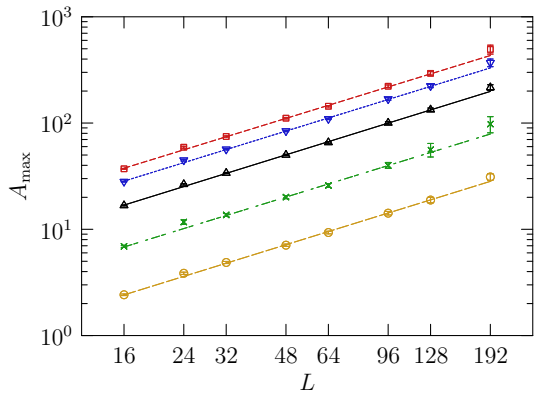}  
  \caption{
    (Color online)
    Fits of the functional forms (\ref{eq:cumulant_scaling}) resp.\
    (\ref{eq:logmagnderiv_scaling}) to the maxima data $A_\mathrm{max}$ of the
      following quantities computed for the case of the 2D Ising model:
      $A = \frac{\d\ln\l|m|^3\r}{\d\beta}$,
    $\frac{\d\ln\l|m|^2\r}{\d\beta}$, $\frac{\d\ln\l|m|\r}{\d\beta}$,  $\frac{\d
      U_4}{\d\beta}$ and  $\frac{\d U_2}{\d\beta}$ (from top to bottom).
    For performing the fits, the correction terms in
    brackets of Eqs.~(\ref{eq:cumulant_scaling}) and (\ref{eq:logmagnderiv_scaling})
    were neglected. The actual fits have been performed on the size range $32\le L\le
    192$. The slopes of the lines are identical to the inverse of the corresponding
    estimates of the correlation length exponent $\nu$ listed in
    Table~\ref{tab:2d_nu}.
}
  \label{fig:nu_fits}
\end{figure}

Although the outlined scheme of Monte Carlo data analysis is completely general, it
is useful to see how it works out for a specific example. In particular, one would
like to know if the typical cross correlations are sufficiently strong to have
significant impact on the results. To answer this question, we performed a
finite-size scaling (FSS) analysis of the ordering transition of the ferromagnetic
Ising model in two and three dimensions.

\subsection{Simulation details}

We studied the critical behavior of the nearest-neighbor, zero-field, ferromagnetic
Ising model with Hamiltonian
\begin{equation}
  \label{eq:ising_hamiltonian}
  {\cal H} = -J\sum_{\l i,j\r}s_i s_j,\;\;\;s_i = \pm 1
\end{equation}
on square and simple cubic lattices of edge length $L$, using periodic boundary
conditions. Close to criticality, an importance-sampling Monte Carlo simulation with
local update rule suffers from critical slowing down, $\tau\sim L^z$, with a
dynamical critical exponent $z \approx 2$. To alleviate this problem, we used the
single-cluster update algorithm \cite{wolff:89a} resulting in a dramatic speed-up of
the relaxation process. For two and three dimensions we performed simulations at a
fixed temperature close to the asymptotic critical temperature for a number of
different system sizes to enable a systematic FSS study. The raw data consisted of
time series with $4\times 10^5$ approximately independent samples of the
configurational energy and magnetization for each system size under
consideration. Using the jackknifing analysis described above, these original time
series were then analyzed using $n$ effectively uncorrelated bins, where $n=100$ was
chosen unless stated otherwise.

\begin{table*}
  \centering
  \begin{minipage}{\textwidth}
  \caption{
    Fit parameters and correlation data
      for estimating the critical exponent
    $\nu$ from single-cluster update Monte Carlo simulations of the 2D Ising model.
    The exponent estimates are extracted from fits of the functional
    forms (\ref{eq:cumulant_scaling}) and (\ref{eq:logmagnderiv_scaling}) to the
    data, neglecting the correction terms in the brackets.
    Deviations from the exact value $\nu = 1$ are
    computed relative to $\nu = 1$ ($\Delta_\mathrm{rel}$) and in multiples of the
    estimated errors listed in the column labeled ``$\sigma$'' ($\Delta_\sigma$).
    \label{tab:2d_nu}
  }
  \begin{ruledtabular}
  \begin{tabular}{crrt{4}t{4}t{2}@{\% }t{2}t{2}rt{4}t{4}t{4}t{4}t{4}}
    \multicolumn{9}{c}{fits} & \multicolumn{5}{c}{correlation coefficients/weights}\\
    & \multicolumn{1}{c}{$L_\mathrm{min}$} & \multicolumn{1}{c}{$L_\mathrm{max}$} &
    \multicolumn{1}{c}{$\nu$} & \multicolumn{1}{c}{$\sigma$} &
    \multicolumn{1}{c}{$\Delta_\mathrm{rel}$} &
    \multicolumn{1}{c}{$\Delta_\sigma$} & \multicolumn{1}{c}{$Q$} &
    \multicolumn{1}{c}{d.o.f.}& \multicolumn{1}{c}{$\displaystyle \frac{\d\ln\langle |m|\rangle}{\d\beta}$}
    & \multicolumn{1}{c}{$\displaystyle \frac{\d\ln\langle m^2\rangle}{\d\beta}$} &
    \multicolumn{1}{c}{$\displaystyle \frac{\d\ln\langle |m|^3\rangle}{\d\beta}$} &
    \multicolumn{1}{c}{$\displaystyle \frac{\d U_2}{\d\beta}$} & \multicolumn{1}{c}{$\displaystyle \frac{\d U_4}{\d\beta}$} \\ \hline
    $\displaystyle \frac{\d\ln\langle |m|\rangle}{\d\beta}$ & 32 & 192 & 1.0085 & 0.0183 & 0.85 & 0.47 & 0.52 & 4 & 1.0000 & 0.9743 & 0.9385 & 0.9197 & 0.8971 \\
    $\displaystyle \frac{\d\ln\langle m^2\rangle}{\d\beta}$ & 32 & 192 & 1.0128 & 0.0194 & 1.28 & 0.66 & 0.47 & 4 & 0.9743 & 1.0000 & 0.9910 & 0.8167 & 0.8687 \\
    $\displaystyle \frac{\d\ln\langle |m|^3\rangle}{\d\beta}$ & 32 & 192 & 1.0175 & 0.0201 & 1.75 & 0.87 & 0.40 & 4 & 0.9385 & 0.9910 & 1.0000 & 0.7431 & 0.8198 \\
    $\displaystyle \frac{\d U_2}{\d\beta}$ & 32 & 192 & 1.0098 & 0.0281 & 0.98 & 0.35 & 0.57 & 4 & 0.9197 & 0.8167 & 0.7431 & 1.0000 & 0.8596 \\
    $\displaystyle \frac{\d U_4}{\d\beta}$ & 32 & 192 & 1.0149 & 0.0511 & 1.49 & 0.29 & 0.70 & 4 & 0.8971 & 0.8687 & 0.8198 & 0.8596 & 1.0000 \\
    \hline\multicolumn{2}{c}{$\bar{\nu}_\mathrm{plain}$} & \multicolumn{1}{l}{$\sigma_\mathrm{uncorr}$} & 1.0127 & 0.0141 & 1.27 & 0.90 & \multicolumn{2}{c}{}& 1.0000& 1.0000& 1.0000& 1.0000& 1.0000\\
    \multicolumn{2}{c}{} &\multicolumn{1}{l}{$\sigma_\mathrm{corr}$} & & 0.0269 & 1.27 & 0.47 & \multicolumn{2}{c}{}& & & & & \\
    \multicolumn{2}{c}{$\bar{\nu}_\mathrm{err}$} & \multicolumn{1}{l}{$\sigma_\mathrm{uncorr}$} & 1.0123 & 0.0102 & 1.23 & 1.21 & \multicolumn{2}{c}{}& 0.3145& 0.2714& 0.2483&0.1322\footnotemark[1]&  0.0336\\
    \multicolumn{2}{c}{} &\multicolumn{1}{l}{$\sigma_\mathrm{corr}$} & & 0.0208 & 1.23 & 0.59 & \multicolumn{2}{c}{}& & & & & \\
    \multicolumn{2}{c}{$\bar{\nu}_\mathrm{cov}$} &
    \multicolumn{1}{l}{$\sigma_\mathrm{corr}$} & 0.9935 & 0.0078 & -0.65 & -0.84 & \multicolumn{2}{c}{}& 5.0067& -2.4259& -0.2807 & -1.1958 &  -0.1043\\
  \end{tabular}
\end{ruledtabular}
\footnotetext[1]{Note that the similar
    Table II of Ref.~\cite{weigel:09} contains a mistake in the last two lines,
    where the weights  $0.1322$ and $0.0336$ as well as  $-1.1958$ and $-0.1043$
    appear interchanged with respect to the correct data represented here.}
\end{minipage}
\end{table*}

\subsection{Finite-size scaling analysis}

There is now a broad consensus that finite-size effects are (in most cases) not
merely a drawback of approaches depending on finite system sizes, but can be turned
into a powerful tool for extracting the asymptotic behavior \cite{barber:domb}. A
number of different practical implementations of this idea in terms of specific FSS
schemes have been derived and successfully applied to the analysis of critical
phenomena, see, e.g.,
Refs.~\cite{ferrenberg:91a,ballesteros:96,hasenbusch:99}. Although our considerations
regarding the data analysis apply rather generally to all these techniques, for
illustrative purposes we concentrate here on the rather popular method outlined in
Ref.~\cite{ferrenberg:91a}. It is focused on the analysis of the locations and values
of extrema of standard thermodynamic quantities such as the specific heat, magnetic
susceptibility, cumulants etc. According to the theory of finite-size scaling
\cite{barber:domb,privman:privman}, the locations of such pseudo-critical points are
shifted away from the true critical coupling $\beta_c$ according to
\begin{equation}
  \label{eq:shift_exponent}
  \beta(A_\mathrm{max},L) = \beta_c + A_0 L^{-\lambda}(1+A_cL^{-w}+\cdots),
\end{equation}
where $A$ denotes an observable with a pseudo-critical maximum such as the specific
heat (for $\alpha \ge 0$). The generic value for the {\em shift exponent\/} $\lambda$
predicted by FSS theory is $\lambda = 1/\nu$, where $\nu$ is the correlation length
exponent (for exceptions see, e.g., Ref.~\cite{janke:02a}). A drawback of using
Eq.~(\ref{eq:shift_exponent}) directly is that non-linear fits in the three
parameters $\beta_c$, $A_0$ and $\lambda$ resp.\ $\nu$ are required even for the
simplest case of ignoring the correction-to-scaling terms in brackets. To alleviate
this problem, it has been suggested to consider quantities such as the magnetization
cumulants \cite{binder:81}
\begin{equation}
  \label{eq:cumulants}
  U_{2i} = 1-\frac{\l |m|^{2i}\r}{3\l|m|^i\r^2},\;\;\;i=1,2,3,\ldots
\end{equation}
for which the maxima of the temperature derivatives have a critical scaling form
\begin{equation}
  \label{eq:cumulant_scaling}
  \left.\frac{\d U_{2i}}{\d\beta}\right|_\mathrm{max} = U_{i,0}L^{1/\nu}(1+U_{i,c}L^{-w}+\cdots),
\end{equation}
and hence allow to determine $\nu$ without prior knowledge of the transition coupling
$\beta_c$. If, again, the correction terms in brackets are ignored, this form even
represents a {\em linear\/} fit (in logarithmic representation)
resulting in very stable results. While initially only the fourth-order cumulant
$U_4$ was considered, the authors of Ref.~\cite{ferrenberg:91a} suggested to use a
variety of different cumulants $U_{2i}$ with $i=1$, $2$, $3$, $\ldots$ to improve the
accuracy of the $\nu$ estimate. A number of further quantities with the same scaling
behavior can be constructed, for instance logarithmic temperature derivatives of the
magnetization,
\begin{equation}
  \label{eq:logmagnderiv_scaling}
  \left.\frac{\d \ln \langle |m|^i\rangle}{\d\beta}\right|_\mathrm{max} =
  D_{i,0}L^{1/\nu}(1+D_{i,c}L^{-w}+\cdots),
\end{equation}
which for $i=1$, $2$, $\ldots$ yields another series of $\nu$ estimates.

Once $\nu$ has been determined from fits of the functional forms
(\ref{eq:cumulant_scaling}) and (\ref{eq:logmagnderiv_scaling}) to the data, one
might return to the shift relation (\ref{eq:shift_exponent}) and (assuming $\lambda =
1/\nu$) determine the transition coupling $\beta_c$ from linear fits with a fixed
value of $\nu$. Finally, the remaining standard critical exponents can be estimated
from the well-known FSS forms of the specific heat $c_V$, the magnetization $m$ and
the magnetic susceptibility $\chi$,
\begin{equation}
  \label{eq:other_exponents}
  \begin{split}
    \left.c_V\right|_\mathrm{max} & = c_0L^{\alpha/\nu}(1+c_kL^{-w}+\cdots),\\
    \l\left|m\right|\r_\mathrm{inf} & = m_0L^{-\beta/\nu}(1+m_kL^{-w}+\cdots),\\
    \left.\chi\right|_\mathrm{max} & = \chi_0L^{\gamma/\nu}(1+\chi_kL^{-w}+\cdots),
  \end{split}
\end{equation}
where $\l\left|m\right|\r_\mathrm{inf}$ denotes the (modulus of the) magnetization at its
inflection point. The directly estimated exponents are therefore $\nu$ and the FSS
exponents $\alpha/\nu$, $\beta/\nu$ and $\gamma/\nu$, which can be combined to yield
$\alpha$, $\beta$, $\gamma$ and $\nu$. The remaining exponents $\delta$ and $\eta$
are not directly determined here; instead we assume that their values are deduced
from the exponents $\alpha$, $\beta$, $\gamma$ and $\nu$ via standard scaling
relations.

\begin{table*}
  \centering
  \caption{
    Fitting and averaging results for estimating the critical coupling $\beta_c$
    of the 2D Ising model from the shifts of pseudo-critical temperatures according
    to Eq.~(\ref{eq:shift_exponent}). For performing the fits, the correlation length
    exponent was fixed at its exact value $\nu = 1$. The column $\Delta_\mathrm{rel}$
    indicates the relative deviation of the estimates from the exact result
    $\beta_c = \frac{1}{2}\ln(1+\sqrt{2})\approx 0.4406868$.
    \label{tab:2d_betac}
  }
  \begin{ruledtabular}
  \begin{tabular}{llt{6}t{6}t{4}@{\% }t{2}t{4}t{4}t{4}t{4}t{4}t{4}t{4}t{4}}
    & \multicolumn{5}{c}{fits} & \multicolumn{8}{c}{correlation coefficients/weights}\\
    && \multicolumn{1}{c}{$\beta_c$} & \multicolumn{1}{c}{$\sigma$} & \multicolumn{1}{c}{$\Delta_\mathrm{rel}$}
    & \multicolumn{1}{c}{$Q$} & \multicolumn{1}{c}{$\displaystyle c_V$}
    & \multicolumn{1}{c}{$\displaystyle \frac{\d\l|m|\r}{\d\beta}$} & \multicolumn{1}{c}{$\displaystyle \frac{\d\ln\langle |m|\rangle}{\d\beta}$}
    & \multicolumn{1}{c}{$\displaystyle \frac{\d\ln\langle m^2\rangle}{\d\beta}$}
    & \multicolumn{1}{c}{$\displaystyle \frac{\d\ln\langle m^3\rangle}{\d\beta}$}
    & \multicolumn{1}{c}{$\displaystyle \frac{\d U_2}{\d\beta}$} & \multicolumn{1}{c}{$\displaystyle \frac{\d U_4}{\d\beta}$}
    & \multicolumn{1}{c}{$\displaystyle \chi$} \\ \colrule
    \multicolumn{2}{c}{$\displaystyle c_V$} & 0.440709 & 0.000101 & 0.0051 & 0.35 & 1.0000 & 0.7881 & 0.3965 & 0.3645 & 0.3569 &  0.3502 & 0.2599 & 0.1394 \\
    \multicolumn{2}{c}{$\displaystyle \frac{\d\l|m|\r}{\d\beta}$} & 0.440798 & 0.000073 & 0.0251 & 0.09 & 0.7881 & 1.0000 & 0.7116 & 0.6417 & 0.6043 & 0.7166 & 0.5345 & 0.6236\\
    \multicolumn{2}{c}{$\displaystyle \frac{\d\ln\langle |m|\rangle}{\d\beta}$} & 0.440711 & 0.000408 & 0.0055 & 0.56 & 0.3965 & 0.7116 & 1.0000 & 0.9740 & 0.9365 & 0.9122 & 0.8613 & 0.6477 \\
    \multicolumn{2}{c}{$\displaystyle \frac{\d\ln\langle m^2\rangle}{\d\beta}$} & 0.440799 & 0.000504 & 0.0254 & 0.42 & 0.3645 & 0.6417 & 0.9740 & 1.0000 & 0.9899 & 0.8119 & 0.8111 & 0.5264 \\
    \multicolumn{2}{c}{$\displaystyle \frac{\d\ln\langle m^3\rangle}{\d\beta}$} & 0.440918 & 0.000567 & 0.0525 & 0.29 & 0.3569 & 0.6043 & 0.9365 & 0.9899 & 1.0000 & 0.7415 & 0.7608 & 0.4554 \\
    \multicolumn{2}{c}{$\displaystyle \frac{\d U_2}{\d\beta}$} & 0.440576 & 0.000338 & -0.0251 & 0.70 & 0.3502 & 0.7166 & 0.9122 & 0.8119 & 0.7415 & 1.0000 & 0.8854 & 0.8413 \\
    \multicolumn{2}{c}{$\displaystyle \frac{\d U_4}{\d\beta}$} & 0.440308 & 0.000708 & -0.0860 & 0.94 & 0.2599 & 0.5345 & 0.8613 & 0.8111 & 0.7608 & 0.8854 & 1.0000 & 0.6265 \\
    \multicolumn{2}{c}{$\displaystyle \chi$} & 0.440699 & 0.000045 & 0.0028 & 0.67 &
    0.1394 & 0.6236 & 0.6477 & 0.5264 & 0.4554 & 0.8413 & 0.6265 & 1.0000 \\\colrule
    $\bar{\beta}_{c,\mathrm{plain}}$ & $\sigma_\mathrm{uncorr}$ & 0.440690 & 0.000151 & 0.0007 & \multicolumn{1}{c}{}& 1.0000& 1.0000& 1.0000& 1.0000& 1.0000& 1.0000& 1.0000& 1.0000
    \\
    & $\sigma_\mathrm{corr}$ & & 0.000322 & 0.0007 & \multicolumn{1}{c}{}& & & & & & & & \\
    $\bar{\beta}_{c,\mathrm{err}}$ & $\sigma_\mathrm{uncorr}$ & 0.440725 & 0.000036 & 0.0086 & \multicolumn{1}{c}{}& 0.1364& 0.2530& 0.0073& 0.0049& 0.0039& 0.0105 & 0.0024 & 0.5815 \\
    & $\sigma_\mathrm{corr}$ & & 0.000059 & 0.0086 & \multicolumn{1}{c}{}& & & & & & & & \\
    $\bar{\beta}_{c,\mathrm{cov}}$ & $\sigma_\mathrm{corr}$  & 0.440687 & 0.000018 & 0.0001 & \multicolumn{1}{c}{}& 0.2601& -0.3347& 0.2086& -0.0566& -0.0186& -0.2997 & 0.0276& 1.2133\\
  \end{tabular}
  \end{ruledtabular}
\end{table*}

For determining the (location and value of the) maxima occurring in
Eqs.~(\ref{eq:shift_exponent}), (\ref{eq:cumulant_scaling}),
(\ref{eq:logmagnderiv_scaling}) and (\ref{eq:other_exponents}), we used the
reweighting technique outlined in Sec.~\ref{sec:histo} starting from the data of a
single simulation per system size performed at or close to the asymptotic transition
point. The derivatives with respect to $\beta$ in Eqs.~(\ref{eq:cumulant_scaling})
and (\ref{eq:logmagnderiv_scaling}) are easily shown to be equivalent to combinations
of moments of energy and magnetization at a single, fixed temperature
\cite{ferrenberg:91a}. For the fourth-order cumulant $U_{4}$, for instance, one has
\begin{equation}
  \label{eq:cumulant_moment}
  \frac{\d U_4}{\d\beta} = \frac{2\l m^4\r[\l m^2\r\l e\r-\l m^2e\r]-\l m^2\r[\l m^4\r\l e\r-\l m^4
    e\r]}{3\l m^2\r^3}.
\end{equation}
Therefore no numerical differentiation is required when making use of the relations
(\ref{eq:cumulant_scaling}) and (\ref{eq:logmagnderiv_scaling}). In some situations
it might be impractical to store the whole time series of original measurements of
internal energy and magnetization, in which case the exact reweighting relation
(\ref{eq:temperature_reweighting}) might be replaced by a Taylor expansion with
respect to $\beta$ around the simulation coupling $\beta_0$, where then cumulants of
$e$ and $m$ appear as the expansion coefficients. In most cases, however, it is much
simpler and more versatile in terms of the data analysis to work with the original
time series. In view of the typically available storage resources today, this
approach should be comfortably feasible in most situations.

Considering the set of critical exponents $\alpha$, $\beta$, $\gamma$, $\nu$ (as well
as $\eta$ and $\delta$), it is useful to recall that they are subject to a number of
exact scaling relations, namely the Rushbrooke identity $\alpha + 2\beta + \gamma =
2$, Fisher's scaling law $\gamma = \nu(2-\eta)$, the relation $\alpha+\beta(1+\delta)
= 2$, as well as (in most cases) the hyperscaling relation $\alpha = 2-\d\nu$, where
$d$ is the spatial dimension \cite{fisher:98}. As a consequence of these four
equations, only two of the six original exponents are independent. While the scaling
relations have been occasionally used to check a set of independently estimated
exponents for consistency, we would like to point out that the existence of these
exact relations should rather be used for {\em improving the precision\/} of exponent
estimates. In particular, in the language of the renormalization group, it is natural
to express the conventional scaling exponents in terms of the {\em scaling
  dimensions\/} $x_t$ and $x_h$ of the operators coupling to temperature and magnetic
field \cite{wj:03a}, respectively, which are the only relevant operators for the
Ising model \cite{henkel:book}. For the four exponents considered here, this means
that
\begin{equation}
  \label{eq:scaling_dimensions}
  \begin{array}{rcl@{\hspace{0.75cm}}rcl}
    \ds\frac{\alpha}{\nu} &=& d-2x_t, &
    \ds\frac{\beta}{\nu} &=& x_h,\\[1ex]
    \ds\frac{\gamma}{\nu} &=& d-2x_h,&
    \ds\frac{1}{\nu} &=& d-x_t,
  \end{array}
\end{equation}
such that $x_h$ can be independently estimated from $x_h = \beta/\nu$ and $x_h =
d/2-\gamma/2\nu$, whereas $x_t$ might be determined from $x_t = d-1/\nu$ as well as
$x_t = d/2-\alpha/2\nu$.

\subsection{Two-dimensional Ising model}

For the two-dimensional (2D) Ising model, single-cluster update simulations were
performed for square lattices of size $L=16$, $24$, $32$, $48$, $64$, $96$, $128$ and
$192$. All simulations were done directly at the asymptotic critical coupling
$\beta_c = \frac{1}{2}\ln(1+\sqrt{2})\approx 0.4406868$. For the range of system
sizes under consideration, it turned out that simulations at this single temperature
were sufficient for reliably studying the pseudo-critical points defined by the
maxima of the various quantities under consideration by reweighting, i.e., the
overlap of histograms between the simulation and analysis temperatures turned out to
be sufficiently large.

We first extracted a number of estimates of the correlation length exponent $\nu$
from investigating the maxima of logarithmic derivatives of magnetization moments for
$i = 1$, $2$ and $3$ as well as the maxima of the derivatives of the second-order and
fourth-order cumulants $U_2$ and $U_4$ using the reweighting scheme outlined
above. The locations and values of the maxima themselves were determined by a golden
section search algorithm \cite{numrec}. The resulting maxima as a function of system
size are shown in Fig.~\ref{fig:nu_fits} together with fits of the forms
(\ref{eq:cumulant_scaling}) and (\ref{eq:logmagnderiv_scaling}) to the data. Here, we
used fits {\em without\/} the correction terms in the brackets of
Eqs.~(\ref{eq:cumulant_scaling}) and (\ref{eq:logmagnderiv_scaling}) on the fit range
$L\ge 32$, which works very well as is apparent from the presentation in
Fig.~\ref{fig:nu_fits} and the corresponding values of the quality-of-fit parameter
$Q$ \cite{brandt:book} listed in the eighth column of Table \ref{tab:2d_nu}. The
fourth and fifth column contain the resulting estimates of the exponent $\nu$
together with the statistical errors estimated from a weighted least-squares fitting
procedure \cite{numrec}. A glance at Table \ref{tab:2d_nu} reveals that all single
estimates are statistically consistent with the exact result $\nu = 1$, but they
exhibit a rather large variation in statistical accuracy with the biggest statistical
error being almost three times larger than the smallest. We use the jackknife
estimator (\ref{eq:jackknife_covariance}) and a second-order jackknifing procedure to
estimate the statistical correlations of the individual estimates of $\nu$. The data
on the right hand side of Table \ref{tab:2d_nu} showing the correlation coefficients
$\rho = \Gamma_{ij}/\sigma_i\sigma_j$ for the different estimates reveal that
correlations between all pairs of estimates are large with $\rho\gtrsim 0.8$. With
all estimates being derived from similar expressions containing magnetic moments,
this result probably does not come as a surprise.

\begin{table}
  \centering
  \caption{
    Averaging results for estimates of the critical exponent $\nu$ and the critical
    coupling $\beta_c$ of the 2D Ising model from non-linear three-parameter fits to
    of the functional form (\ref{eq:shift_exponent}). The observables used are those
    listed in Table~\ref{tab:2d_betac}.
    \label{tab:three_parameter}
  }
  \begin{ruledtabular}
    \begin{tabular}{clt{4}t{4}t{6}t{6}}
      & & \multicolumn{1}{c}{$\nu$} & \multicolumn{1}{c}{$\sigma$} &
      \multicolumn{1}{c}{$\beta_c$} & \multicolumn{1}{c}{$\sigma$} \\ \colrule
      $\bar{\theta}_\mathrm{plain}$ & $\sigma_\mathrm{uncorr}$ & 0.8101 & 0.0428 & 0.439491 & 0.000337 \\
                          & $\sigma_\mathrm{corr}$   &        & 0.0973 &          & 0.000715 \\
      $\bar{\theta}_\mathrm{err}$ & $\sigma_\mathrm{uncorr}$ & 0.8949 & 0.0228 & 0.440295 & 0.000099 \\
                          & $\sigma_\mathrm{corr}$   &        & 0.0435 &          & 0.000169 \\
      $\bar{\theta}_\mathrm{cov}$ & $\sigma_\mathrm{corr}$ & 0.9980 & 0.0148 &
      0.440658 & 0.000072 \\ \colrule
      exact               &                         & 1.0000 &        & 0.440687 &         \\
    \end{tabular}
  \end{ruledtabular}
\end{table}

Under these circumstances, one might wonder whether it is worthwhile to attempt a
linear combination of the form (\ref{eq:linear_combination}) of the various $\nu$
estimates rather than quoting the single most precise estimate as final result, which
in the present case is given by the value $\nu = 1.0085(183)$ resulting from the FSS
of $\d\ln\l|m|\r/\d\beta$. For the purpose of combining estimates, we consider the
traditional approaches of taking a {\em plain average\/} $\bar{\nu}_\mathrm{plain}$
with
\begin{equation}
  \alpha_i^\mathrm{plain} = \frac{1}{k}  
\end{equation}
as well as the {\em error-weighted average\/} $\bar{\nu}_\mathrm{err}$ of
Eq.~(\ref{eq:error_weighted}) and compare them to the truly optimal {\em
  covariance-weighted average\/} $\bar{\nu}_\mathrm{cov}$ defined by the weights of
Eq.~(\ref{eq:covariance_weighted}). Ignoring the presence of correlations (as was
the case in most previous studies), one would estimate the error associated to the
plain average as
\begin{equation}
  \label{eq:uncorr_plain_error}
  \sigma^{2\,\mathrm{plain}}_\mathrm{uncorr} = \frac{1}{k^2}\sum_i \sigma^2(\hat{\nu}_i),
\end{equation}
and, likewise, the variance of the error-weighted average is given by
$\sigma^{2\,\mathrm{err}}_\mathrm{uncorr}$ as defined in
Eq.~(\ref{eq:uncorrelated_variance}). The {\em true\/} variances of
$\bar{\nu}_\mathrm{plain}$ and $\bar{\nu}_\mathrm{err}$ in the presence of
correlations, on the other hand, can also be easily derived formally, and will
contain the elements of the covariance matrix $\Gamma$ of the individual estimates
$\hat{\nu}_i$. From the practical perspective, the jackknifing analysis outlined here
{\em automatically\/} takes those correlations into account. We refer to these
correctly defined variances with the notation $\sigma^2_\mathrm{corr}$.

The plain, error-weighted and covariance-weighted averages for $\nu$ with the
corresponding variance estimates are listed in the lower part of columns four and
five of Table \ref{tab:2d_nu}. As with the individual estimates, each of the three
averages is statistically compatible with the exact result $\nu = 1$. While the
na\"{\i}ve error estimates $\sigma_\mathrm{uncorr}$ seem to indicate that performing
the plain or error-weighted average reduces statistical fluctuations compared to the
single estimates, taking correlations into account with the jackknifing scheme
resulting in $\sigma^2_\mathrm{corr}$ reveals that variances are grossly underestimated
by $\sigma^2_\mathrm{uncorr}$ and, in fact, compared to both the plain
($\sigma_\mathrm{corr} = 0.0269$) and error-weighted ($\sigma_\mathrm{corr} =
0.0208$) averages the single estimate of $\nu$ stemming from $\d\ln\l|m|\r/\d\beta$
has smaller statistical fluctuations ($\sigma = 0.0183$). Performing those averages
therefore {\em decreases precision\/} instead of improving it! The truly optimal
average of Eq.~(\ref{eq:covariance_weighted}), on the other hand, results in the
estimate $\nu = 0.9935(78)$, whose fluctuation is about 2--3 times smaller than those
of the error-weighted average and the single most precise estimate. The reduced
variance of this last estimate seems to be corroborated by the smallest deviation
also from the exact result $\nu = 1$. A glance at the data collected in Table
\ref{tab:2d_nu} reveals that, somewhat astonishingly, the optimal average is smaller
than {\em all\/} of the individual estimates of $\nu$ (see also the graphical
representation of this fact in Fig.~1 of Ref.~\cite{weigel:09}). This situation
which, of course, can never occur for the error-weighted average where all weights
$0\le \alpha_i\le 1$, is connected to the fact that the more general weights of
Eq.~(\ref{eq:covariance_weighted}) are unbounded and, in particular, can become
negative. This fact reflects in the computed weights for the different averaging
schemes collected in the lower right hand part of Table \ref{tab:2d_nu}. Clearly, the
weights for the error-weighted and covariance-weighted averages are dramatically
different and, in particular, some of the latter turn out to be negative. It is
intuitively clear that such negative weights are necessary to cancel the effects of
strong mutual correlations. The asymmetry in the weights leading to the possibility
of the average lying outside of the range of the individual estimates results from
the asymmetry of the individual variances in connection with the cross
correlations. This effect can be explicitly understood for the case of only two
estimates, cf.\ the discussion in Appendix \ref{sec:app}.

One might wonder whether the suggested weighting scheme requiring to estimate the
full covariance matrix is statistically robust. It is clear, for instance, that the
jackknife estimator (\ref{eq:jackknife_covariance}) for the covariance will become
more precise as more jackknife blocks are used --- at the expense of an increased
computational effort. To check for such effects we repeated our analysis while using
$n=200$ instead of $n=100$ jackknife blocks. Most of the estimates for the
correlation coefficients are almost unchanged by this new analysis with the largest
deviation being of the order of 3\%. The same holds true for the resulting weights in
the optimal average, where only the weight of the estimate resulting from
$\d\ln\l|m|^3\r/\d\beta$ changes substantially from $\alpha = -0.2807$ to $\alpha =
-0.5499$. The final optimal estimate $\bar{\nu} = 0.9908(78)$ is fully compatible
statistically with the analysis using $100$ jackknife blocks. Using (as a consistency
check) completely independent simulations for producing the individual estimates of
$\nu$, on the other hand, indeed results in a unit matrix of correlation coefficients
within statistical errors and, consequently, the error-weighted and
covariance-weighted averages coincide in this limit. Finally, we also find that the
numerical inversion of the covariance matrix required for computing the weights in
Eq.~(\ref{eq:covariance_weighted}) is in general stable and unproblematic. It is
clear, however, that in the presence of very strong correlations the resulting
weights of individual estimates will depend sensitively on the entries of the
covariance matrix, since in the limit of {\em perfect\/} correlations all choices of
weights become degenerate, see also the discussion of the case of only two estimates
in Appendix \ref{sec:app}.

We now turn to the determination of the transition coupling $\beta_c$ from the shift
relation (\ref{eq:shift_exponent}). We considered the locations of the extrema of the
specific heat $c_V$, the slope $\d\l|m|\r/\d\beta$ of the (modulus of the)
magnetization, the logarithmic derivatives $\d\ln \langle |m|^i\rangle/\d\beta$ for
$i=1$, $2$ and $3$, the cumulant derivatives $\d U_2/\d\beta$ and $\d U_4/\d\beta$ as
well as the magnetic susceptibility $\chi$. In order to most clearly demonstrate the
effects of the present correlations, we first performed fits of the form
(\ref{eq:shift_exponent}) using the exact correlation length exponent $\nu = 1$. The
corresponding fit results are collected in Table \ref{tab:2d_betac}. For the fits we
ignored the correction terms indicated in the brackets of
Eq.~(\ref{eq:shift_exponent}), leaving out the smallest system sizes instead. This
approach appears justified in view of the good fit qualities reflected in the $Q$
values of Table \ref{tab:2d_betac}. As for the fits for determining $\nu$, all single
estimates of $\beta_c$ are consistent with the true asymptotic values of $\beta_c
\approx 0.4406868$ within error bars. The corresponding standard deviations, however,
vary dramatically, decreasing by a factor of $15$ from the estimate resulting from
$\d U_4/\d\beta$ to that of the susceptibility $\chi$. The results of the correlation
analysis are presented on the right-hand side of Table \ref{tab:2d_betac}: while the
logarithmic magnetization derivatives and cumulants again show very strong
correlations, the results of the remaining quantities are somewhat more independent,
showing, in particular, a rather clear separation of the energetic from the magnetic
sector. For the averages of single estimates the present correlations again lead to a
significant underestimation of the true variance for the plain and error-weighted
cases and, in fact, both of them are {\em less\/} precise than the best single
estimate stemming from the scaling of the susceptibility $\chi$, cf.\ the data in the
lower part of Table \ref{tab:2d_betac}. The truly optimal average of
Eq.~(\ref{eq:covariance_weighted}) results in $\beta_c = 0.440687(18)$, where the
statistical error is about threefold reduced compared to the error-weighting
scheme. Very similar results are found when using the value $\nu = 0.9935(78)$ found
from the analysis summarized in Table \ref{tab:2d_nu}, where we arrive at a final
covariance-weighted average of $\beta_c = 0.440658(17)[35]$. Here, the second error
estimate in square brackets refers to the sensitivity of the result for $\beta_c$ to
the uncertainty in $\nu$ indicated above, which turns out to be symmetric with
respect to upwards and downwards deviations of $\nu$ here.

As an alternative to the two-step process of first determining $\nu$ from the
relations (\ref{eq:cumulant_scaling}) and (\ref{eq:logmagnderiv_scaling}) and only
afterwards estimating $\beta_c$ from Eq.~(\ref{eq:shift_exponent}), one might
consider direct fits of the form (\ref{eq:shift_exponent}) to the maxima data of the
8 observables listed above determining $\nu$ and $\beta_c$ in one go. Here, again,
fits on the range $32 \le L \le 192$ neglecting any corrections to the leading
scaling behavior are found to be sufficient. The results for the plain,
error-weighted and covariance-weighted averages for both parameters, $\nu$ and
$\beta_c$, are collected in Table \ref{tab:three_parameter}. Consistent with the
previous results, it is seen that neglecting correlations in error estimation leads
to a sizable underestimation of errors and, on the other hand, using the optimal
weighting scheme of Eq.~(\ref{eq:covariance_weighted}) statistical errors are
significantly reduced, an effect which is also nicely illustrated by the very good
fit of the resulting parameter estimates with the exact values.

\begin{table}
  \centering
  \caption{
    Determining the magnetic and energetic scaling dimensions $x_h$ and $x_t$ 
    of the 2D Ising model by weighted averages over various
    individual estimates.
    \label{tab:scaling_dimension}
  }
  \begin{ruledtabular}
    \begin{tabular}{clt{4}t{4}t{4}t{4}}
      & & \multicolumn{1}{c}{$x_h$} & \multicolumn{1}{c}{$\sigma$} &
      \multicolumn{1}{c}{$x_t$} & \multicolumn{1}{c}{$\sigma$} \\ \colrule
      $\bar{\theta}_\mathrm{plain}$ & $\sigma_\mathrm{uncorr}$ & 0.1219 & 0.0027 & 1.0085 & 0.0117 \\
                                   & $\sigma_\mathrm{corr}$   &        & 0.0021 &          & 0.0213 \\
      $\bar{\theta}_\mathrm{err}$ & $\sigma_\mathrm{uncorr}$   & 0.1261 & 0.0016 & 1.0048  & 0.0082 \\
                                   & $\sigma_\mathrm{corr}$   &        & 0.0013 &          & 0.0136 \\
      $\bar{\theta}_\mathrm{cov}$ & $\sigma_\mathrm{corr}$   & 0.1250 & 0.0010 &
      1.0030 & 0.0096 \\ \colrule
      exact                        &                         & 0.1250 &        & 1.0000 &         \\
    \end{tabular}
  \end{ruledtabular}
\end{table}

\begin{table*}
  \centering
  \caption{
    Fit parameters and correlation data for estimating the critical exponent
    $\nu$ from single-cluster update Monte Carlo simulations of the 3D Ising model.
    Fits of the functional form (\ref{eq:logmagnderiv_scaling}) including the
    correction term were used for the logarithmic magnetization derivatives
    $\d\ln\l|m|^i\r/\d\beta$ for $i=1$, $2$ and $3$, while fits of the form
    (\ref{eq:cumulant_scaling}) {\em without\/} correction term were used for
    the derivatives of the cumulants $U_2$ and $U_4$. The relevant reference values is $\nu = 0.6301(4)$
    taken from Ref.~\cite{pelissetto:02}.
    \label{tab:3d_nu}
  }
  \begin{ruledtabular}
  \begin{tabular}{crrt{4}t{4}t{2}@{\% }t{2}t{2}rt{4}t{4}t{4}t{4}t{4}}
    \multicolumn{9}{c}{fits} & \multicolumn{5}{c}{correlation coefficients/weights}\\
    & \multicolumn{1}{c}{$L_\mathrm{min}$} & \multicolumn{1}{c}{$L_\mathrm{max}$} &
    \multicolumn{1}{c}{$\nu$} & \multicolumn{1}{c}{$\sigma$} &
    \multicolumn{1}{c}{$\Delta_\mathrm{rel}$} &
    \multicolumn{1}{c}{$\Delta_\sigma$} & \multicolumn{1}{c}{$Q$} &
    \multicolumn{1}{c}{d.o.f.}& \multicolumn{1}{c}{$\displaystyle \frac{\d\ln\langle |m|\rangle}{\d\beta}$}
    & \multicolumn{1}{c}{$\displaystyle \frac{\d\ln\langle m^2\rangle}{\d\beta}$} &
    \multicolumn{1}{c}{$\displaystyle \frac{\d\ln\langle |m|^3\rangle}{\d\beta}$} &
    \multicolumn{1}{c}{$\displaystyle \frac{\d U_2}{\d\beta}$} & \multicolumn{1}{c}{$\displaystyle \frac{\d U_4}{\d\beta}$} \\ \hline
    $\displaystyle \frac{\d\ln\langle |m|\rangle}{\d\beta}$ & 8 & 128 & 0.6358 & 0.0127 & 0.91 & 0.45 & 0.61 & 5 & 1.0000 & 0.9809 & 0.9490 & 0.4401 & 0.4507 \\
    $\displaystyle \frac{\d\ln\langle m^2\rangle}{\d\beta}$ & 8 & 128 & 0.6340 & 0.0086 & 0.63 & 0.46 & 0.71 & 5 & 0.9809 & 1.0000 & 0.9910 & 0.4357 & 0.4630 \\
    $\displaystyle \frac{\d\ln\langle |m|^3\rangle}{\d\beta}$ & 8 & 128 & 0.6326 & 0.0062 & 0.39 & 0.40 & 0.77 & 5 & 0.9490 & 0.9910 & 1.0000 & 0.4363 & 0.4639 \\
    $\displaystyle \frac{\d U_2}{\d\beta}$ & 32 & 128 & 0.6313 & 0.0020 & 0.20 & 0.62 & 0.54 & 3 & 0.4401 & 0.4357 & 0.4363 & 1.0000 & 0.9267 \\
    $\displaystyle \frac{\d U_4}{\d\beta}$ & 32 & 128 & 0.6330 & 0.0024 & 0.46 & 1.20 & 0.77 & 3 & 0.4507 & 0.4630 & 0.4639 & 0.9267 & 1.0000 \\
    \hline\multicolumn{2}{c}{$\bar{\nu}_\mathrm{plain}$} & \multicolumn{1}{l}{$\sigma_\mathrm{uncorr}$} & 0.6334 & 0.0038 & 0.52 & 0.85 & \multicolumn{2}{c}{}& 1.0000& 1.0000& 1.0000& 1.0000& 1.0000\\
    \multicolumn{2}{c}{} &\multicolumn{1}{l}{$\sigma_\mathrm{corr}$} & & 0.0067 & 0.52 & 0.49 & \multicolumn{2}{c}{}& & & & & \\
    \multicolumn{2}{c}{$\bar{\nu}_\mathrm{err}$} &
    \multicolumn{1}{l}{$\sigma_\mathrm{uncorr}$} & 0.6322 & 0.0015 & 0.33 & 1.35 & \multicolumn{2}{c}{}& 0.0106& 0.0254& 0.0503& 0.5315 & 0.3823\\
    \multicolumn{2}{c}{} &\multicolumn{1}{l}{$\sigma_\mathrm{corr}$} & & 0.0024 & 0.33 & 0.84 & \multicolumn{2}{c}{}& & & & & \\
    \multicolumn{2}{c}{$\bar{\nu}_\mathrm{cov}$} &
    \multicolumn{1}{l}{$\sigma_\mathrm{corr}$} & 0.6300 & 0.0017 & -0.01 & -0.05 &
    \multicolumn{2}{c}{}& 0.2485& -1.5805& 1.6625& 0.7948 & -0.1253 \\
  \end{tabular}
\end{ruledtabular}
\end{table*}

Finally, we turn to the determination of the remaining critical exponents. As
outlined above, we do this by combining different estimates using covariance analysis
to improve the results for the scaling dimensions, thus ensuring that the scaling
relations are fulfilled exactly. From a glance at Eq.~(\ref{eq:scaling_dimensions})
one reads off that the magnetic scaling dimension $x_h$ can be determined from $x_h =
\beta/\nu$ and $x_h = d/2-\gamma/2\nu$. We therefore determine $\beta/\nu$ from the
FSS of the (modulus of the) magnetization at its inflection point and estimate
$\gamma/\nu$ from the FSS of the susceptibility maxima, resulting in $\beta/\nu =
0.1167(54)$ and $\gamma/\nu = 1.7458(40)$, respectively. As the correlation analysis
reveals, the two resulting estimates of $x_h$ are {\em anti-}correlated to a
considerable degree with correlation coefficient $-0.64$. As a consequence,
conventional error analysis neglecting correlations {\em over-}estimates statistical
fluctuations. Still, choosing optimal weights according to
Eq.~(\ref{eq:covariance_weighted}) is able to reduce variance, resulting in a
combined estimate $x_h = 0.1250(10)$ right on top of the exact result $x_h = 1/8$,
cf.\ the data collected in Table \ref{tab:scaling_dimension}. The energetic scaling
dimension $x_t$, on the other hand, might be computed from $x_t = d-1/\nu$ as well as
$x_t = d/2-\alpha/2\nu$. We therefore use the five individual estimates of $\nu$
listed in Table \ref{tab:2d_nu} as well as the FSS of the maximum of the specific
heat to estimate $x_t$. The latter fits are somewhat problematic due to the
logarithmic singularity of the specific heat corresponding to $\alpha/\nu = 0$, and
it turns out that a fit of the form
$$
c_{V,\mathrm{max}} = c_0+c_1L^{\alpha/\nu}\ln L
$$
including a scaling correction is necessary to describe the data. Combining all
individual estimates in an optimal way, we arrive at $x_t = 1.0030(96)$, well in
agreement with the exact result $x_t = 1$, cf.\ the right hand side of Table
\ref{tab:scaling_dimension}.

\subsection{Three-dimensional Ising model}

Cluster-update simulations of the ferromagnetic Ising model in three dimensions (3D)
were performed for simple-cubic lattices of edge lengths $L=8$, $12$, $16$, $24$,
$32$, $48$, $64$, $96$ and $128$. All simulations were performed at the coupling
$\beta = 0.221\,654\,9$ reported in a high-precision study as estimate for the
transition point \cite{bloete:99a}, since it turned out that the maxima of the
various quantities under consideration were all within the reweighting range of this
chosen simulation point for the system sizes and lengths of time series at hand.

For determining the correlation-length exponent $\nu$ we again considered the scaling
of the logarithmic magnetization derivatives $\d\ln\l|m|^i\r/\d\beta$ for $i=1$, $2$
and $3$ and the derivatices of the cumulants $U_2$ and $U_4$. We find scaling
corrections to be somewhat more pronounced than for the two-dimensional model for the
system sizes studied here. For the logarithmic magnetization derivatives we therefore
performed fits of the form (\ref{eq:logmagnderiv_scaling}) including the correction
term on the full range $8\le L\le 128$, where the resulting values of the effective
correction exponent $w$ were $w=0.57(63)$ ($i=1$), $w=0.69(56)$ ($i=2$) and
$w=0.80(52)$ ($i=3$), respectively. For the cumulants $U_2$ and $U_4$, on the other
hand, corrections were too small to be fitted reliably with our data, such that they
were effectively taken into account by dropping the small lattice sizes instead,
while using fits of the form (\ref{eq:cumulant_scaling}) with $U_{i,c}=0$ fixed. The
corresponding fit data are collected in Table \ref{tab:3d_nu}. The estimated standard
deviations of the individual estimates are again found to be very heterogeneous, but
the correlations between the different estimates are somewhat smaller than in two
dimensions, in particular between the magnetization derivatives and the cumulants,
cf.\ Table \ref{tab:3d_nu}. Comparing to the case of fits without corrections, it is
seen that this latter effect is partially due to the use of two different fit forms
for the two types of quantities. (The fits for $U_2$ and $U_4$ also include a reduced
range of lattice sizes which could lead to a decorrelation, but this effect is found
to be much less important than the difference in the fit forms.) Considering the
averages of individual estimates, as a result of these smaller correlations the
underestimation of statistical errors in the na\"{\i}ve approach as well as the
reduction of variance through the optimized estimator (\ref{eq:covariance_weighted})
is somewhat less dramatic than for the two-dimensional model, but the qualitative
behavior appears to be very much the same. As our final estimate we quote $\nu =
0.6300(17)$, very well in agreement with the reference value $\nu = 0.6301(4)$ taken
from a survey of recent literature estimates compiled in Ref.~\cite{pelissetto:02}.

In a second step we determined the transition coupling from fits of the functional
form (\ref{eq:shift_exponent}) to the maxima of the quantities listed in Table
\ref{tab:2d_betac}. As for the $\nu$ fits, however, the inclusion of an effective
correction term as indicated in Eq.~(\ref{eq:shift_exponent}) turned out to be
necessary for a faithful description of the scaling data. The plain, error-weighted
and covariance-weighted averages of the corresponding estimates are listed in the
first two data columns of Table \ref{tab:various_exponents_3d} together with their
standard deviations, the results being consistent with the reference value. We also
tried non-linear three-parameter fits of the form (\ref{eq:shift_exponent}) to the
data, determining $\nu$ and $\beta_c$ simultaneously. For this case, the precision of
the data is not high enough to reliably include corrections to scaling. Still, the
improved results are well consistent with the reference values of
Refs.~\cite{bloete:99a,pelissetto:02}, cf.\ the middle columns of Table
\ref{tab:various_exponents_3d}.
\begin{table*}
  \centering
  \caption{
    Different averages for the 3D Ising model
    and the associated standard deviations for the transition
    coupling $\beta_c$ from fits of the form (\ref{eq:shift_exponent}) with $\nu =
    0.6301$ fixed, from non-linear three-parameter fits of the form
    (\ref{eq:shift_exponent}) yielding $\nu$ and $\beta_c$ simultaneously,
    and for the magnetic and
    energetic scaling dimensions according to Eq.~(\ref{eq:scaling_dimensions}). The
    reference values for $x_h$ and $x_t$ have been computed from the values
    $\beta = 0.3265(3)$ and $\nu = 0.6301(4)$ taken from Ref.~\cite{pelissetto:02}
    via Eq.~(\ref{eq:scaling_dimensions}).
    \label{tab:various_exponents_3d}
  }
  \begin{ruledtabular}
    \begin{tabular}{clt{8}t{8}t{4}t{4}t{7}t{7}t{4}t{4}t{4}t{4}}
      & & \multicolumn{2}{c}{Eq.~(\ref{eq:shift_exponent}), $\nu = 0.6301$} &
      \multicolumn{4}{c}{Eq.~(\ref{eq:shift_exponent})} &
      \multicolumn{4}{c}{Eq.~(\ref{eq:scaling_dimensions})} \\
      & & \multicolumn{1}{c}{$\beta_c$} & \multicolumn{1}{c}{$\sigma$} &
      \multicolumn{1}{c}{$\nu$} & \multicolumn{1}{c}{$\sigma$} &
      \multicolumn{1}{c}{$\beta_c$} & \multicolumn{1}{c}{$\sigma$} &
      \multicolumn{1}{c}{$x_h$} & \multicolumn{1}{c}{$\sigma$} &
      \multicolumn{1}{c}{$x_t$} & \multicolumn{1}{c}{$\sigma$} \\ \colrule
      $\bar{\theta}_\mathrm{plain}$ & $\sigma_\mathrm{uncorr}$ & 0.22165681 &
      0.00000108 & 0.6020 & 0.0105 & 0.2216530 & 0.0000025 & 0.51364 & 0.00401 &
      1.4137 & 0.0138 \\
                                   & $\sigma_\mathrm{corr}$   &            &
                                   0.00000170 &        & 0.0150 & & 0.0000032 & &
                                   0.00435 & & 0.0184\\
      $\bar{\theta}_\mathrm{err}$ & $\sigma_\mathrm{uncorr}$   & 0.22165741 &
      0.00000059 & 0.6247 & 0.0062 & 0.2216550 & 0.0000008 & 0.51489 & 0.00381 &
      1.4180 & 0.0038\\
                                   & $\sigma_\mathrm{corr}$   &            &
                                   0.00000114 &        & 0.0077 & & 0.0000016 & &
                                   0.00413 & & 0.0061\\
      $\bar{\theta}_\mathrm{cov}$ & $\sigma_\mathrm{corr}$     & 0.22165703 &
      0.00000085 & 0.6381 & 0.0044 & 0.2216552 & 0.0000011 & 0.51516 & 0.00412 &
      1.4121 & 0.0043\\ \colrule
      reference                        &                     & 0.22165459 &
      0.00000006 & 0.6301 & 0.0004 & 0.22165459 & 0.00000006 & 0.51817 & 0.00058 &
      1.4130 & 0.0010\\
    \end{tabular}
  \end{ruledtabular}
\end{table*}

Finally, we also considered the scaling dimensions $x_h$ and $x_t$. For the magnetic
scaling dimension, we find that the determinations from $x_h = \beta/\nu$ and $x_h =
3/2-\gamma/2\nu$ are only very weakly correlated, such that the error-weighted and
covariance-weighted averages are very similar, see the right hand side of Table
\ref{tab:various_exponents_3d}. Larger correlations are present again between the
different estimates of the energetic scaling dimension $x_t$ from the various
estimates of $\nu$ via $x_t = 3-1/\nu$ and the scaling of the specific heat via $x_t
= 3/2-\alpha/2\nu$, leading to a considerable improvement in precision of the
optimal average over the plain and error-weighting schemes. The results for both
scaling dimensions are well compatible with the values $x_h = 0.51817(58)$ and $x_t =
1.4130(10)$ extracted from the reference values of Ref.~\cite{pelissetto:02}.

\section{Conclusions}

Time series data from Markov chain Monte Carlo simulations are usually analyzed in a
variety of ways to extract estimates for the parameters of interest such as, e.g.,
critical exponents, transition temperatures, latent heats etc. As long as at least
some of these estimates are based on the same simulation data, a certain degree of
cross correlations between estimators is unavoidable. We have shown for the case of a
finite-size scaling analysis of the ferromagnetic nearest-neighbor Ising model on
square and cubic lattices that more often than not, such correlations are very
strong, with correlation coefficients well above 0.8. While such correlations,
although their existence is rather obvious, have been traditionally mostly neglected
even in high-precision numerical simulation studies, it was shown here that their
presence is of importance at different steps of the process of data analysis, and
neglecting them leads to systematically wrong estimates of statistical fluctuations
as well as non-optimal combination of single estimates into final averages.

As far as the general statistical analysis of simulation data is concerned, it has
been discussed that traditional prescriptions such as error propagation have their
shortcomings, in particular as soon as non-parametric steps such as the determination
of a maximum via reweighting or fitting procedures come into play. These problems are
circumvented by resorting to the class of non-parametric resampling schemes, of which
we have discussed the jackknife technique as a conceptually and practically very
simple representative. Using this technique, we have outlined a very general
framework of data analysis for MCMC simulations consisting of (a) a transformation
of the original set of time series into an auxiliary set of ``binned'' series, where
successive samples are approximately uncorrelated in time and (b) a general
jackknifing framework, where the required steps of computing a parameter estimate ---
possibly including reweighting or fitting procedures etc. --- are performed on the
full underlying time series apart from a small window cut out from the data stream
allowing for a reliable and robust estimate of variances and covariances as well as
bias effects without any non-stochastic approximations. While this technique of data
analysis is not new, we feel that it still has not found the widespread use it
deserves and hope that the gentle and detailed introduction given above will
contribute to a broader adoption of this approach.

A particular example of where the presence of cross correlations comes into play
occurs when taking averages of different estimates for a parameter from the same data
base. Neglecting correlations there leads to (a) systematically wrong, most often too
small, estimates of statistical errors of the resulting averages and (b) a
sub-optimal weighting of individual values in the average leading to
larger-than-necessary variances. Correct variances can be estimated straightforwardly
from the jackknifing approach, while optimal weighting involves knowledge of the
covariance matrix which is a natural byproduct of the jackknife technique as well. We
have discussed these concepts in some detail for the case of a finite-size scaling
analysis of the critical points of the 2D and 3D Ising models. It is seen there that
the plain and error-weighted averages most oftenly used in fact can have larger
fluctuations than the most precise single estimates entering them, but this flaw is
not being detected by the conventional analysis due to the generic underestimation of
variances. On the contrary, by using the truly optimal weighting of individual
estimates an often substantial reduction of statistical fluctuations as compared to
the error-weighting scheme can be achieved. For some of the considered examples, a
threefold reduction in standard deviation, corresponding to saving an about tenfold
increase in computer time necessary to achieve the same result with the conventional
analysis, can be achieved with essentially no computational overhead. In view of
these results, heuristic rules such as, e.g., taking an error-weighted average using
the smallest single standard deviation as an error estimate are clearly found to be
inadequate. We therefore see only two statistically acceptable ways of dealing with
the existence of several estimates for the same quantity: (a) select the single most
precise estimate and discard the rest or (b) combine all estimates in a statistically
optimal way taking cross correlations into account. Needless to say, the latter
approach is generally preferable in that it leads to more precise results at very low
costs.

We suggest to use the existence of scaling relations between the critical exponents
for the case of a continuous phase transition to improve the precision of estimates
by considering the scaling dimensions as the parameters of primary
interest. Performing the corresponding analysis taking cross correlations into
account, results in a set of critical exponents with reduced statistical
fluctuations that fulfill the scaling relations exactly. An application of this type
of approach initially suggested in Ref.~\cite{weigel:09} for using mean-value
relations such as Callen identities or Schwinger-Dyson equations instead of scaling
relations has been discussed in Ref.~\cite{fernandez:09}.

While the examples discussed were specific, it should be clear that the method
itself is rather generic, and should apply to all data sets generated from MCMC
simulations. In particular, it is easy to envisage applications in the theory of
critical phenomena, reaching from classical statistical mechanics \cite{binder:book}
over soft matter physics \cite{holm:05} to quantum phase transitions
\cite{vojta:03b}, or for studying first-order phase transitions \cite{janke:03}. The
range of applications is not restricted to MCMC simulations, however, but applies
with little or no modifications to other random sampling problems, such as, e.g.\
stochastic ground-state computations \cite{weigel:06b,weigel:06c} or the sampling of
polymer configurations with chain-growth methods \cite{grassberger:97a,bachmann:03a}.

\begin{acknowledgments}
  M.W.\ acknowledges support by the DFG through the Emmy Noether Programme under
  contract No.\ WE4425/1-1 as well as computer time provided by NIC J\"ulich under
  grant No.\ hmz18.
\end{acknowledgments}

\appendix

\section{Optimal average of two correlated variables\label{sec:app}}

\begin{figure}[tb]
  \centering
  \includegraphics[keepaspectratio=true,scale=0.75,trim=75 48 75 78]{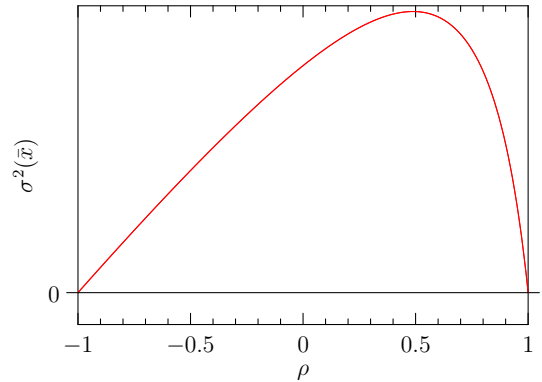}
  \caption
  {
    (Color online)
    Generic form of the minimal variance $\sigma^2(\bar{x})$ of
    Eq.~(\ref{eq:app_optimal_variance}) as a function of the correlation coefficient
    $\rho$.
    \label{fig:two_variable}
  }
\end{figure}

Consider a general average of two random variables $x_1$ and $x_2$ \cite{janke:97},
$$
\bar{x} = \kappa x_1+(1-\kappa)x_2,
$$
where $0\le\kappa\le 1$. According to Eq.~(\ref{eq:variance_of_average}), the
variance of $\bar{x}$ is
\begin{equation}
  \label{eq:app_variance}
  \sigma^2(\bar{x}) = \kappa^2\sigma_1^2+2\kappa(1-\kappa)\rho\sigma_1\sigma_2
  +(1-\kappa)^2\sigma_2^2,  
\end{equation}
where $\sigma_1^2$ and $\sigma_2^2$ are the variances of $x_1$ and $x_2$,
respectively, and $\rho$ denotes the correlation coefficient of $x_1$ and $x_2$,
$\rho =\Gamma_{12}/\sigma_1\sigma_2$. Eq.~(\ref{eq:app_variance}) is a quadratic
form in $\kappa$, which has a minimum as long as
$$
\sigma_1^2+\sigma_2^2-2\rho\sigma_1\sigma_2 > 0,
$$
which is almost always fulfilled since $-1\le\rho\le 1$:
$$
\sigma_1^2+\sigma_2^2-2\rho\sigma_1\sigma_2\ge (\sigma_1-\sigma_2)^2 \ge 0.
$$
Equality holds only for $\sigma_1 = \sigma_2 = \sigma$ and $\rho = 1$, in which case
{\em any\/} choice of $\kappa$ yields the same variance $\sigma^2(\bar{x}) =
\sigma^2$. In all other cases, the optimal weights are
\begin{equation}
  \begin{split}
    \kappa & =
    \frac{1/\sigma_2^2-\rho/\sigma_1\sigma_2}{1/\sigma_1^2+1/\sigma_2^2-2\rho/\sigma_1\sigma_2},\\
    1-\kappa & = \frac{1/\sigma_1^2-\rho/(\sigma_1\sigma_2)}{1/\sigma_1^2+1/\sigma_2^2-2\rho/(\sigma_1\sigma_2)},  
  \end{split}
  \label{eq:app_kappa}
\end{equation}
and the resulting variance of the average is
\begin{equation}
  \label{eq:app_optimal_variance}
  \sigma^2(\bar{x}) = \frac{1-\rho^2}{1/\sigma_1^2+1/\sigma_2^2-2\rho/(\sigma_1\sigma_2)}.  
\end{equation}
A number of observations are immediate
\begin{itemize}
\item[(i)] For the uncorrelated case $\rho = 0$, one arrives back at the
  error-weighted average of Eqs.~(\ref{eq:error_weighted}) and
  (\ref{eq:uncorrelated_variance}).
\item[(ii)] In the correlated case, and for fixed variances $\sigma_1^2$ and
  $\sigma_2^2$, the variance $\sigma^2(\bar{x})$ smoothly depends on the correlation
  coefficient $\rho$. It has maxima at $\sigma_1/\sigma_2$ and $\sigma_2/\sigma_1$,
  only one of which is in the range $|\rho| \le 1$. Notably, the relevant maximum is
  always at non-negative values of $\rho$.
\item[(iii)] For $\rho = \pm 1$, the variance {\em vanishes identically\/}, apart
  from the singular case $\sigma_1 = \sigma_2$ and $\rho = 1$.
\end{itemize}
The generic form of $\sigma^2(\bar{x})$ as a function of $\rho$ is depicted in
Fig.~\ref{fig:two_variable}. In the presence of moderate correlations, therefore,
{\em anti-correlations\/} are preferable over correlations in terms of reducing the
variance of the average. Note that the result (\ref{eq:app_optimal_variance}) is
different from that of Eq.~(8) in Ref.~\cite{fernandez:09}, since the definition of
correlation coefficient used there is different from that in our situation of taking
an average. Instead of measuring the correlation between $x_1$ and $x_2$, their
definition refers to the correlation of $x_1$ and $x_2-x_1$.

\begin{figure}[b]
  \centering
  \includegraphics[keepaspectratio=true,scale=0.75]{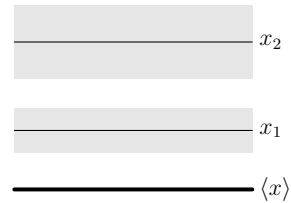}
  \caption
  {
    For strong positive correlations, i.e., for $\rho > \sigma_1/\sigma_2$ in the
    case $\sigma_1 < \sigma_2$, the most likely location of the true expectation $\l
    x\r$ is outside of the bracket $[\min(x_1,x_2),\max(x_1,x_2)]$.
    \label{fig:fluctuation}
  }
\end{figure}

The weights $\kappa$ and $1-\kappa$ of Eq.~(\ref{eq:app_kappa}) are not restricted to
be between zero and one. It is easy to see that for $\kappa > 1$ or $\kappa < 0$, the
average $\bar{x}$ is in fact outside of the bracket
$[\min(x_1,x_2),\max(x_1,x_2)]$. This seemingly paradoxical effect is easily
understood from the optimal weights derived here. From Eq.~(\ref{eq:app_kappa}) one
reads off that the weights $\kappa$ and $1-\kappa$ leave the range $0\le \kappa,
1-\kappa\le 1$ as soon as $\rho\ge \sigma_1/\sigma_2$ resp.\ $\rho\ge
\sigma_2/\sigma_1$, depending on whether $\sigma_1 < \sigma_2$ or $\sigma_2 <
\sigma_1$, that is, only for strong positive correlations to the right of the maximum
in Fig.~\ref{fig:two_variable}. Thus, if the smaller of $x_1$ and $x_2$ has the
smaller variance (and both are strongly correlated), the average is below both
values. If the larger value has the smaller variance, the optimal average is above
both values. The asymmetry comes here from the difference in variance. To understand
this intuitively, assume for instance that $x_1 < x_2$ and $\sigma_1 < \sigma_2$ with
strong positive correlations $\rho > \sigma_1/\sigma_2$. It is most likely, then,
that $x_1$ and $x_2$ deviate in the {\em same\/} direction from the true mean $\l
x\r$. Since $\sigma_1 < \sigma_2$, the deviation of $x_1$ should be generically
smaller than that of $x_2$. For $x_1 < x_2$, however, this is only possible if $\l
x\r < x_1 < x_2$. This is illustrated in Fig.~\ref{fig:fluctuation}.

%\bibliography{citeulike}

\end{document}